\documentclass[traditabstract]{aa}

\usepackage{graphicx}
\usepackage{txfonts}
\usepackage{amssymb}

\usepackage{natbib}
\bibpunct{(}{)}{;}{a}{}{,}

\newcommand{\beq}{\begin{equation}}
\newcommand{\eeq}{\end{equation}}
\newcommand{\bea}{\begin{eqnarray}}
\newcommand{\eea}{\end{eqnarray}}
\newcommand{\req}[1]{Eq.~(\ref{#1})}

\newcommand{\dd}{\mathrm{d}} 
\newcommand{\gcc}{\mbox{g cm$^{-3}$}}
\newcommand{\mel}{m} 
\newcommand{\mpr}{M} 
\newcommand{\am}{a_\mathrm{m}}
\newcommand{\bete}{\beta_\mathrm{e}}
\newcommand{\betp}{\beta_\mathrm{p}}
\newcommand{\Ea}{E_\mathrm{a}^\infty}
\newcommand{\He}{H_\mathrm{e}}
\newcommand{\Hp}{H_\mathrm{p}}
\newcommand{\kc}{K_\mathrm{cr}}
\newcommand{\kp}{K_\perp}
\newcommand{\mH}{m_\mathrm{H}}
\newcommand{\mion}{M}
\newcommand{\mred}{m_\ast}
\newcommand{\omc}{\omega_{\mathrm{c}}} 
\newcommand{\omce}{\omega_{\mathrm{c}}} 
\newcommand{\omcp}{\Omega_{\mathrm{c}}} 
\newcommand{\omci}{\Omega_{\mathrm{c}}} 
\newcommand{\Teff}{T_\mathrm{eff}}
\newcommand{\Teffbb}{T_\mathrm{eff,bb}^\infty}
\newcommand{\ASR}[1]{{Adv.\ Space Res.}, {#1}}
\newcommand{\AstL}[1]{{Astron.\ Lett.}, {#1}}
\newcommand{\FCP}[1]{{Fundam.\ Cosmic Phys.}, {#1}}
\newcommand{\JPB}[1]{{J.\ Phys.\ B}, {#1}}
\newcommand{\NuCi}[1]{{Nuovo Cimento}, {#1}}
\newcommand{\PRv}[1]{{Phys.\ Rev.}, {#1}}
\newcommand{\RMP}[1]{{Rev.\ Mod.\ Phys.}, {#1}}
\newcommand{\RPP}[1]{{Rep.\ Prog.\ Phys.}, {#1}}
\newcommand{\SovJETP}[1]{{Sov.\ Phys. JETP}, {#1}}

\begin{document}

\title{Cyclotron harmonics in opacities 
of isolated neutron star atmospheres}
                                                                                
\author{A. Y. Potekhin
  \inst{1,2,3}}
\mail{\email{palex@astro.ioffe.ru}}

\institute{CRAL (UMR CNRS No. 5574), 
Ecole Normale Sup\'{e}rieure de Lyon,
69364 Lyon Cedex 07, France
\and
Ioffe Physical-Technical Institute,
Politekhnicheskaya 26, 194021 St.~Petersburg, Russia
\and
Isaac Newton Institute of Chile, 
         St.~Petersburg Branch, Russia}

\date{Received 13 April 2010 / Accepted 6 May 2010}


\abstract{Some X-ray dim isolated neutron stars (XDINS) and central
compact objects in supernova remnants (CCO) contain absorption features in
their thermal soft X-ray spectra. It has been hypothesized
that this absorption may relate to periodic peaks in
free-free absorption opacities,  caused by either Landau quantization
of  electron motion in magnetic fields $B\lesssim10^{11}$~G or analogous
quantization of ion motion in magnetic fields $B>10^{13}$~G. 
Here, I review
the physics behind cyclotron quantum harmonics in free-free 
photoabsorption, discuss different approximations for their calculation,
and  explain why the  ion cyclotron harmonics (beyond the fundamental)
cannot be observed.
}

\maketitle

\section{Introduction} 
\label{sect:intro}

Thermal radiation from neutron stars can provide important
information about their physical properties. Among neutron
stars with thermal-like radiation spectra (see, e.g., reviews by 
\citealt{kaspietal06} and \citealt{zavlin09}), there are two classes of
objects of particular interest: central compact objects (CCOs;
see, e.g., \citealt{deluca08}) in supernova remnants and
X-ray dim isolated neutron stars (XDINSs, or the Magnificent Seven;
see, e.g., review by \citealt{turolla09}). 

The CCOs are young, radio-quiet isolated neutron stars with
relatively weak magnetic fields $B \sim(10^{10}-10^{11})$~G
(e.g., \citealt{halperngotthelf10} and references therein).
The XDINSs are older  and are believed to have much stronger
fields $B \gtrsim(10^{12}-10^{13})$~G
\citep{haberl07,vankerkwijkkaplan07,turolla09}. For some CCOs and
XDINSs, there are estimates of $B$, and in some cases only upper
limits to $B$ are available.

In the past decade, broad absorption lines have been detected in
the thermal spectra of several isolated neutron stars (see, e.g., 
\citealt{vankerkwijk04,haberl07,vankerkwijkkaplan07}, and
references therein). In all but one case, the energies $\Ea$ of the
absorption are centered on the range 0.2--0.7 keV and the 
effective black-body temperatures are
$\Teffbb\approx0.1$ keV.
Here and hereafter,
the Boltzmann constant is suppressed, and
the superscript ``$\infty$'' indicates a
redshifted value. 
In particular, it has been found that:
(i)
 the spectrum of \object{RX J1605.3+3249} with
$\Teffbb\approx96$ eV has a broad absorption at
$\Ea\approx0.4$--0.5 keV and a possible second absorption at 0.55
keV \citep{vankerkwijketal04,vankerkwijk04};
(ii)
 \object{RX J0720.4-3125}
  exhibits an absorption feature at $\Ea\approx0.27$ keV
\citep{haberletal04} and a possible second absorption at 0.57 keV
\citep{hambaryanetal09}, while the
effective black-body temperature varies over years across the range
$\Teffbb\approx86$--95 eV \citep{hohleetal09};
(iii)
 the spectrum of \object{RBS1223} (RX J$1308.6+2127$) 
 was reproduced by a model
with $\Teffbb\approx102\pm2$ eV and two absorption lines at
$\Ea\sim0.3$ keV and $\Ea\sim0.6$ keV \citep{schwopeetal07};
and (iv)
 the
spectrum of \object{RBS1774} (1RXS J$214303.7+065419$)
with  $\Teffbb\approx102$~eV shows
indications of a
line at $\Ea\approx0.3$--0.4 keV and an absorption edge at 
$0.73$--0.75 keV
\citep{cropperetal07,kaplanvankerkwijk09,schwopeetal09}.
The first discovered isolated 
neutron star with absorption lines,
CCO \object{1E 1207.4-5209},
has two
absorption features centered on $\Ea\approx0.7$
keV and 1.4 keV \citep{sanwaletal02}
and an effective black-body
temperature (which may be nonuniform) of $\Teffbb\sim0.16$--0.32
keV \citep{zavlinetal98,delucaetal04}. For this object,
two more harmonically spaced absorption features (at
$\Ea\approx2.1$ keV and 2.8 keV) were tentatively
detected \citep{bignamietal03,delucaetal04}, but were later
shown to be statistically insignificant \citep{morietal05}. 
We note
that realistic values of the effective temperature
$\Teff^\infty$, obtained using atmosphere models, can differ 
 from $\Teffbb$ by a
factor $\lesssim2$--3 (see, e.g., \citealt{zavlin09},
and references therein).

Many authors 
\citep[e.g.,][]{sanwaletal02,bignamietal03,delucaetal04}
have considered the theoretical possibility that the absorption
lines in the thermal spectra of the CCOs and XDINSs
may be produced by cyclotron harmonics,
formed because
of quantum transitions between different Landau levels of charged
particles in strong magnetic fields. 
\citet{zaneetal01} discussed
this possibility prior to the observational discovery
of these absorption features. 
The fundamental cyclotron
energy equals 
\beq
  \hbar\omce=\hbar
{eB/\mel c}=11.577~B_{12}\textrm{ keV}
\label{omce}
\eeq
 for the electrons and
$\hbar\omci=\hbar{ZeB/ \mion c}=6.35\,(Z/A) B_{12}$~eV for the
ions, where $\mel$ and $\mion$ are the electron and
 ion masses, respectively, $Z$ and  $A$ are the ion
charge and mass numbers, and $B_{12}\equiv B/10^{12}$~G. In the
following, we consider
protons, whose cyclotron energy is
\beq
 \hbar\omcp=\hbar eB/ \mpr c = 6.305\,B_{12} \textrm{ eV}.
\label{omcp}
\eeq

Beginning with the pioneering work of
\citet{GnedinSyunyaev74}, numerous papers have been devoted to the
physics and modeling of cyclotron lines in X-ray spectra of
accreting neutron stars 
\citep[e.g.,][]{DV77,PSY80,WWL93,ArayaHarding,ArayaG-Harding,nishimura05,nishimura08}.
These emission lines have been
observed in many works following their discovery by
\citet{Truemperetal78}. Cyclotron harmonics have been found in 
spectra of several X-ray pulsars in binaries  \citep[e.g.,][and
references therein]{rodes-rocaetal09,enotoetal08,pottschmidtetal04},
and up to four
harmonics were registered for one of them
\citep{santangeloetal99}.

In the photospheres of isolated neutron stars, unlike X-ray binaries,
the typical energies of charged particles are nonrelativistic. In
this case, first-order cyclotron transitions of free charged
particles are dipole-allowed only between neighboring equidistant
Landau levels and form a single cyclotron resonance with no
harmonics. Special relativity and non-dipole  corrections at the
energies of interest can be estimated to be
$\mathrm{max}(\Teff,E_\mathrm{a})/\mel c^2 \sim10^{-3}$ for the
electrons and $\mathrm{max}(\Teff,E_\mathrm{a})/\mpr c^2<10^{-6}$
for the protons.

Beyond the first order in interactions,  transitions between
distant Landau states are also allowed in the nonrelativistic theory.
They are, in particular, caused by Coulomb interactions
between plasma particles. Thus cyclotron harmonics appear in
free-free (bremsstrahlung) cross-sections.
To obtain $E_\mathrm{a}\sim0.1$--1 keV, one may assume
either the electron cyclotron harmonics at $B\sim
10^{10}$--$10^{11}$~G, according to \req{omce}, or proton
cyclotron harmonics at $B\sim 10^{13}$--$10^{14}$~G, according to
\req{omcp}. 

\citet{pavlovshibanov78} presented the calculations
of spectra for isolated neutron stars with prominent
\emph{electron} cyclotron harmonics due to the free-free
absorption in the atmosphere. 
\citet{SulPavWer} performed a similar atmosphere modeling 
and concluded, in agreement with
\citet{zaneetal01}, that electron cyclotron harmonics could be
observed in CCO spectra.
Proton cyclotron harmonics cannot be calculated
based on the assumption of classical proton motion, used by these authors.

In this paper, I review the physics  of free-free photoabsorption
in strong magnetic fields, discuss restrictions on different
published approximations for free-free opacities, and present
numerical results that demonstrate the relative strengths of the
\emph{electron and proton} cyclotron resonances under the conditions
characteristic of the atmospheres of isolated neutron stars with
strong magnetic fields. This gives a graphic explanation of the
smallness of the ion cyclotron harmonics. I also demonstrate that
the contribution of bound-bound and bound-free transitions
to the opacities of neutron stars with $B>10^{13}$~G
is much larger than that of the proton
cyclotron harmonics.

In Sect.\,\ref{sect:particle}, quantum mechanical integrals of
motion and wave functions of a charged particle in a magnetic
field are recalled for subsequent use. Section \ref{sect:ep} is
devoted to the properties of an electron-proton system in a
magnetic field that is quantizing for both particles:
 general equations for calculation of wave functions are
given, and the Born approximation is considered in detail.
In the same order, general expressions and Born
approximation are considered in Sect.\,\ref{sect:ff} for
photoabsorption matrix elements and cross-sections. Section
\ref{sect:cycl} gives numerical examples of cyclotron harmonics
in free-free photoabsorption with discussion and comparison of
various approximations. Consequences for
the CCOs and XDINSs are discussed in
Sect.\,\ref{sect:discussion}, and Sect.\,\ref{sect:sum} presents our
summary.

\section{Charged particles in a magnetic field}
\label{sect:particle}

Since special relativity effects are of minor importance in
the atmospheres of isolated neutron stars, we use
nonrelativistic quantum mechanics.

We assume that the magnetic field vector $\vec{B}$ is
along the $z$ axis and consider its vector potential
the cylindrical gauge to be
\beq
   \vec{A}(\vec{r}) = \frac12\,\vec{B}\times(\vec{r}-\vec{r}_A)
\label{cyl}
\eeq
with an arbitrary center $\vec{r}_A$ in the $xy$ plane.

We recall the description
of a charged particle
in a uniform magnetic field
\citep[e.g.,][]{johnsonlippmann49,LaLi-QM,johnsonetal83}. 
The Hamiltonian equals the kinetic
energy operator 
\beq
   H^{(1)}=\frac{m\dot{\vec{r}}^2}{2} = H_\perp^{(1)}
      + \frac{p_{z}^2}{2m},
\qquad
     H_\perp^{(1)} = \frac{m\dot{\vec{r}}_{\perp}^2}{2},
\label{H1}
\eeq
where $m$ is the mass,  
\beq
   m \dot{\vec{r}} = \vec{p} - (Q/c)\,\vec{A}(\vec{r})
\label{r-dot}
\eeq
is the kinetic momentum, $Q$ is the charge, and
$\vec{p}$ is the canonical momentum
conjugate to $\vec{r}$. In \req{H1} and hereafter,
``$\perp$'' denotes the ``transverse'' part,
related to the motion in the $xy$ plane.

A classical particle moves along a
spiral around
the normal to the $xy$ plane at the \emph{guiding center}
$\vec{r}_{c}$. In quantum mechanics, $\vec{r}_{c}$ is an operator,
related to the \emph{pseudomomentum} operator
\beq
      \hbar\vec{k} = m\dot{\vec{r}} + (Q/c)\,\vec{B}\times\vec{r},
\label{k_i}
\eeq
where
$\vec{r}_{c}=(c/QB^2)\,\hbar \vec{k}\times\vec{B}$. 
Its cartesian coordinates $(x_c,y_c)$ commute with
$H_\perp^{(1)}$, but do not commute with each other:
$[x_c,y_c]=-\mathrm{i}\hbar c/QB$. 
Another important
integral of motion is the $z$-projection of the angular momentum
$\ell_z=(QB/c)\,(r_c^2/2-H_\perp^{(1)}/m\omc^2)$,
where $\omc=|Q|B/mc$ is the cyclotron frequency.

The eigenvalues of $H_\perp^{(1)}$ are given by 
$E^\perp_n=(n+\frac12)\hbar\omc$,
where $n=0,1,2,\ldots$ is the Landau quantum number. The
simultaneous
eigenvalues of $\ell_z$ are ($\mathrm{sign}\,Q) \,\hbar
s$ with integer $s\geq -n$, and eigenvalues of the squared
guiding center $r_c^2$ equal to $\am^2(2s+2n+1)$, where $\am=(\hbar
c/|Q|B)^{1/2}$ is the so-called magnetic length.

In general, $H_\perp^{(1)}$ should be supplemented by 
$(-\vec{B}\cdot\hat{\vec{\mu}})$, where
$
  \hat{\vec{\mu}}=g_\mathrm{mag}\,(e/2mc)\,\hat{\vec{S}}
$
is the intrinsic
magnetic moment of the particle,
$\hat{\vec{S}}$ is the spin operator, and
$g_\mathrm{mag}$ is the spin $g$-factor
($g_\mathrm{mag}=-2.0023$ and $5.5857$
for the electron and the proton, respectively).
In most applications, one can choose
 the representation where the electron and proton spins
 have definite 
$z$-projections
 $\pm\hbar/2$ 
 and set the electron $g$-factor to $-2$, thus regarding
the excited electron Landau levels as double degenerate.

The form of a wave function depends on a choice of the
gauge for $\vec{A}(\vec{r})$. 
We consider the cylindrical gauge given by \req{cyl} centered
on the coordinate origin ($\vec{r}_A=\vec{0}$). The
eigenfunctions of $H^{(1)}$ and $\ell_z$
in the coordinate representation are
\beq
   \Psi_{n, s,k_z}(\vec{r}) =
   \frac{\mathrm{e}^{\mathrm{i}k_zz}}{
   L_z^{1/2}} \times\left\{
   \begin{array}{ll}\Phi_{n,s}(\vec{r}_\perp),
   &\mbox{if $Q<0$}\\
\Phi^\ast_{n,s}(\vec{r}_\perp),
   &\mbox{if $Q>0$}
   \end{array}\right.
   ,
\label{magnelpsi1}
\eeq
where $k_z=p_z/\hbar$ is the particle wave number along the
field,
$L_z$ is the normalization length, 
$\vec{r}_\perp=(x,y)=(r_\perp\cos\phi,r_\perp\sin\phi)$,
\beq
   \Phi_{n,s}(\vec{r}_\perp) = \frac{\mathrm{e}^{-\mathrm{i}s\phi}}{
   \sqrt{2\pi}\,\am}\,I_{n+s,n}(r_\perp^2/2\am^2)
\label{Phi-Landau}
\end{equation}
is Landau function,
the asterisk denoting a complex conjugate,
and
$I_{n'n}(u)$ 
is a Laguerre function \citep[e.g.,][]{SokTer}.

We define cyclic components of any vector $\vec{a}$ as
$
a_{\pm1}=(a_x\pm\mathrm{i}a_y)/\sqrt{2}
$
and 
$a_0=a_z$.
The transverse cyclic components of the
kinetic momentum operator given by \req{r-dot}
transform one Landau state $|n,s\rangle_\perp$,
characterized by $\Phi_{ns}^{(\ast)}(\vec{r}_\perp)$,
into another Landau state
\beq
   m\dot{\vec{r}}_{\alpha}\, |n,s\rangle_\perp =
     \frac{\hbar\tilde\alpha}{\mathrm{i}\am}
   \sqrt{n+1/2+\tilde\alpha/2}
   \,|n+\tilde\alpha,s-\tilde\alpha\rangle_\perp,
\label{pi1+-}
\eeq
where 
$\alpha=\pm1$,
$\tilde\alpha=\alpha$ if $Q<0$,
and $\tilde\alpha=-\alpha$ if $Q>0$.

\section{Electron-proton system in a magnetic field}
\label{sect:ep}

The Hamiltonian of the electron-proton pair 
(i.e., of H atom) is
\beq
   H = \He^{(1)}+\Hp^{(1)} - 
     \frac{e^2}{r_{\mathrm{ep}}},
\label{Hgeneral}
\eeq
where
$\vec{r}_{\mathrm{ep}}=\vec{r}_\mathrm{e}-\vec{r}_\mathrm{p}$. The kinetic part can be written as
\beq
  \He^{(1)}+\Hp^{(1)} =
  \frac{\mpr\dot{\vec{r}}_\mathrm{p}^2}{2} + \frac{\mel\dot{\vec{r}}_\mathrm{e}^2}{2}
   = \frac{\mH\dot{\vec{R}}^2}{2}  + \frac{\mred\dot{\vec{r}}^2}{2},
\eeq
where $\mel$ and $\mpr$ are the electron and ion masses,
$\vec{R}=(\mpr/\mH)\,\vec{r}_\mathrm{p}+(\mel/\mH)\,\vec{r}_\mathrm{e}$ is the center of
mass,
$\mH=\mel+\mpr$ is the total mass, and $\mred=\mel \mpr/\mH$ is the
reduced mass.

Since the electron and the proton have opposite
charges, their orbiting in the transverse plane is
accompanied by a
drift across the magnetic field lines with 
velocity $\vec{v}_\mathrm{drift}$ depending on
the distance between their guiding centers 
or equivalently on the total pseudomomentum $\vec{K}$
\beq
   \vec{r}_c = \vec{r}_{c,e}-\vec{r}_{c,p}
   = \frac{c}{eB^2}\,\vec{B}\times\vec{K},
\quad
  \vec{K} = 
    \vec{P} -
   \frac{e}{2c}\,\vec{B}\times\vec{r}_{\mathrm{ep}},
\eeq
where
$\vec{P}$ is the canonical
momentum conjugate to $\vec{R}$.

In quantum mechanics, it is not only true that
the pseudomomentum operator
$\vec{K}= \hbar\vec{k}_\mathrm{e} + \hbar\vec{k}_\mathrm{p}$ commutes with $H$,
but also that its cartesian components $(K_x,K_y,K_z)$
commute with each other.
Therefore, all components of $\vec{r}_c$ can be determined
simultaneously. Coordinate eigenfunctions of
the pseudomomentum operator with eigenvectors $\vec{K}$
are given by \citep{GorDzyal}
\beq
   \Psi(\vec{r}_\mathrm{e},\vec{r}_\mathrm{p}) = \exp\left[\frac{\mathrm{i}}{\hbar}
   \left(\vec{K} +
   \frac{e}{c}\,\vec{B}\times\vec{r}_{\mathrm{ep}}\right)\cdot\vec{R} \right]
   \,\psi_{\vec{K}}(\vec{r}_{\mathrm{ep}}).
\label{PsiH}
\eeq
From the general Schr\"odinger equation $H\Psi=E\Psi$,
one can derive an equation for $\psi_{\vec{K}}(\vec{r}_{\mathrm{ep}})$,
which has the form
\beq
   H_\mathrm{rel}\,\psi_{\vec{K}} = E\,\psi_{\vec{K}},
\label{Schrod}
\eeq
where the effective Hamiltonian $H_\mathrm{rel}$ depends on 
$\vec{K}$.

\subsection{Exact solution}
\label{sect:exact}

Solutions of \req{Schrod} for arbitrary $\vec{K}$ in strong
magnetic fields in the cylindrical gauge 
represented by \req{cyl} were
obtained by \citet{VDB92} and \citet{P94} for bound states,
and by \citet{PP97} for
continuum states of the electron-proton system. 
\citet{P94} used the variable 
$\vec{r}=\vec{r}_{\mathrm{ep}}-\vec{r}_B$ as an independent argument
of the wave function
and found that the most convenient
parametrization in \req{cyl} is
$
   \vec{r}_A = [(\mpr - \mel)/\mH]\,\vec{r}_B.
$
Then
\beq
   H_\mathrm{rel} 
   = \frac{P_z^2}{2\mH} + \frac{p_z^2}{2\mred} +
   H_\perp - \frac{e^2}{|\vec{r}+\vec{r}_B|},
\label{Hrel}
\eeq
where
\beq
   H_\perp 
    = \frac{{\pi}_\perp^2}{2\mred} - \frac{e}{\mpr c}
   \,\vec{B}\cdot(\vec{r}\times\vec{p}) + H_K
\label{Hperp}
\eeq
is the Hamiltonian 
of the harmonic motion in the $xy$ plane,
$\vec{p}$ is the momentum conjugate to $\vec{r}$,
\bea
   \vec{\pi} &=& \vec{p} + \frac{e}{2c}\,\vec{B}\times\vec{r},
\label{pi}
\\
   H_K &=& \frac{K_B^2}{2\mH} + \frac{e}{\mH c}
   \,\vec{K}_B\cdot(\vec{B}\times\vec{r}),
\quad\mbox{and}\quad
   \vec{K}_B = \vec{K} + \frac{e}{c}\,\vec{B}\times\vec{r}_B.
\eea
We note that $H_K=0$ when $\vec{r}_B=\vec{r}_c$.

The first term in \req{Hrel} is the total kinetic energy along $z$,
uncoupled from the relative electron-proton motion, therefore
we set $P_z=0$ without loss of generality.

The eigenvalues of $H_\perp$ equal 
$
   E^\perp_{ns} = E^\perp_{n} + E^\perp_{N},
$
where $n\geq0$ and $N=n+s\geq0$ are the electron and proton Landau numbers,
respectively, and $\hbar s$ are eigenvalues
of the relative
angular momentum projection operator
($\ell_{z,p}-\ell_{z,e}$).

We construct numerical solutions of \req{Schrod} 
in the energy representation 
for $\vec{r}_B=\eta\,\vec{r}_c$ ($\eta\in[0,1]$)
in the form 
\beq
   \psi_\kappa^{(\eta)}(\vec{r}) = 
   \sum_{n's'} \Phi_{n's'}(\vec{r}_\perp)\, 
   g^{(\eta)}_{n',s';\,\kappa}(z),
\label{expan}
\eeq
where $\kappa$ is the composite quantum number enumerating
quantum states. 
One retains in \req{expan} as many terms
($n=0,1,2,\ldots,n_\mathrm{max}$;
$s=-n,-n+1,-n+2,\ldots,s_\mathrm{max}$) as needed
to reach the desired accuracy.
We choose a principal (``leading'')
term $(n,s)$ and define ``longitudinal'' energy
of the state $|\kappa\rangle$ as 
$
   E^\|_\kappa = 
   E_\kappa - E^\perp_{ns}.
$
The functions
$g^{(\eta)}_{n',s';\,\kappa}(z)$ are computed
from 
\bea&&
   \Bigg(
     -\frac{\hbar^2}{2\mred}\,\frac{\dd^2}{\dd z^2}
     + V_{n''s'',n''s''}(r_B,z) +\langle n''s''\,|\,H_K\,|\,n''s''
     \rangle_\perp
\nonumber\\&&\quad
     + E^\perp_{n''s''} - E^\perp_{ns} - E^\|_{\kappa}
     \,\Bigg)\,
     g_{n''s'';\,\kappa}(z)=
\nonumber\\&&\qquad
   - {\sum}'\big( V_{n''s'',n's'}(r_B,z) 
      +
 \langle n''s''\,|H_K|\,n's' \rangle_\perp \big)
     \,g_{n's';\,\kappa}(z),
\label{Hsystem}
\eea
where $n''=0,1,2,\ldots,n_\mathrm{max}$,
$s''=-n'',-n''+1,-n''+2,\ldots,s_\mathrm{max},
$
$\Sigma'$ denotes the sum over all pairs $(n',s')$
except $(n'', s'')$,
and
\beq
    V_{n''s'',n's'}(r_B,z) =
     \big\langle n''s''\,\big|\,
     -e^2/|\vec{r}+\vec{r}_B|
     \,\big|\,n's'\big\rangle_\perp
\label{Veff}
\eeq
are effective potentials (see \citealt{P94}
for calculation of
these potentials and
matrix elements
$\langle n''s''\,|\,H_K\,|\,n's' \rangle_\perp$).

\subsubsection{Bound states}

Bound states 
of the H atom can be numbered as
$|\kappa\rangle = |ns\nu\vec{K}\rangle$, where
$\nu=0,1,2,\ldots$ enumerates energy levels 
for every fixed pair ($n,s$)
and controls the $z$-parity according to the relation
$g_{n',s';\,\kappa}(-z)=(-1)^\nu g_{n',s';\,\kappa}(z)$.
The longitudinal energies
$
   E^\|_\kappa \equiv E^\|_{ns\nu}(K)
$
are determined from the system of equations 
in \req{Hsystem}
together with the longitudinal wave functions
$g_{n's',\kappa}(z)$.

The atomic states are qualitatively different for small and large
$\kp$ values. For small $\kp$, the electron remains mostly around
the proton, the energy dependence on $\kp$ is nearly quadratic,
so that the transverse velocity  $v_\mathrm{drift}=\partial E_\kappa/
\partial\kp$ is nearly proportional to $\kp$, i.e.,
$v_\mathrm{drift}\approx\kp/\mH^\perp$. The effective mass
$\mH^\perp$ exceeds $\mH$ and increases with
increasing $B$ \citep{VB88}. For large $\kp$, the atomic state is
\emph{decentered} \citep{GorDzyal}: the electron finds itself
mostly around $\vec{r}_c$, rather than around the proton. In the
latter case, $v_\mathrm{drift}$ decreases with increasing
$\kp$. The two families of states are separated by the critical
value of the pseudomomentum, $\kc \approx (2\mH
|E^\|_{ns\nu}(0)|)^{1/2}$,  where the electron
wave function is mostly asymmetric,
while the transverse velocity of
the atom reaches a maximum \citep{VDB92,P94,P98}.

\subsubsection{Continuum}
\label{sect:WFC}

Wave functions of the continuum are computed using
the same expansion, \req{expan}, and system of equations
in \req{Hsystem}, as for the bound states,  but 
for a given energy $E$ for every $z$-parity. 
The solution is based on a
translation of the usual $R$-matrix formalism
\citep[e.g.,][]{Seaton83} to the case of a strong magnetic field.
Now $|\kappa\rangle = |ns,E,\vec{K},\pm\rangle$, where
``$\pm$'' reflects the symmetry condition
$g(z)=\pm g(-z)$.
Numbers $n$ and $s$
mark a selected \emph{open channel},
defined for 
$E^\|_\kappa \equiv E-E^\perp_{ns} > 0$
by asymptotic conditions
at $z\to+\infty$
\beq
   g_{n_os_o;\,\kappa}^\mathrm{real}(z) \sim
    \delta_{n_on}\delta_{s_os}
   \sin[\phi_{ns}(z)]
   + R_{n_os_o;\,ns}
   \cos[\phi_{n_os_o}(z)],
\label{asymp}
\eeq
where the pairs $(n_o,s_o)$, as well as $(n,s)$,
relate to the open channels
($E>E^\perp_{n_o,s_o}$),
\beq
   \phi_{ns}(z)=k_{ns} z + (\mred e^2/\hbar^2 k_{ns})\,\ln(k_{ns}z)
\label{phase}
\eeq
is the $z$-dependent part of the phase of the wave function
at $z\to+\infty$,
and $k_{ns}=\sqrt{2\mred (E-E^\perp_{ns})}/\hbar$
is the wave number. 
For the \emph{closed channels}, defined by the opposite
inequality $E^\perp_{n_c,s_c}>E$, one should select
$g_{n_cs_c;\,\kappa}(z) \to0$ at $z\to\infty$. If $I_o$ is the total
number of open channels at given $E$, then the set of solutions,
defined by Eqs.~(\ref{Hsystem}) and (\ref{asymp}),
constitute a complete set of $I_o$ independent
real basis functions. 
The quantities $R_{n'_o,s'_o;\,n_os_o}$ constitute the 
reactance matrix $\mathcal{R}$, which has dimensions
$I_o\times I_o$. If the wave functions are normalized 
according to the condition
\beq
   \int_{\mathbb{R}^2}\dd\vec{r}_\perp
   \int_{-L_z/2}^{L_z/2}\dd z \, |\psi_{\vec{K}}(\vec{r}_\perp,z)|^2 = 1,
\label{L_z}
\eeq
then the reactance matrix satisfies the relation
\beq
   k_{n's'} R_{ns;\,n's'} = k_{ns} R_{n's';\,ns},
\eeq
which differs from the usual symmetry relation \citep{Seaton83}.

The representation with $\eta=1$ must be used for
continuum states, to ensure that the
right-hand side (r.h.s.) of \req{Hsystem} vanishes at
$|z|\to\infty$, which is required by the asymptotic condition
of \req{asymp}.

For a final state of a transition,
one should use 
wave functions describing outgoing waves.
The basis of outgoing waves with definite $z$-parity
is defined by the asymptotic
conditions
\beq
   g_{n_os_o;\,ns}^\mathrm{out}(z) \sim
   \delta_{n_on}\delta_{s_os}\,
   \mathrm{e}^{\mathrm{i}\phi_{ns}(z)}
   - S_{n_os_o;\,ns}^*\,
   \mathrm{e}^{-\mathrm{i}\phi_{n_os_o}(z)},
\label{asout}
\eeq
where $S_{n_os_o;ns}$ are the elements of the scattering matrix
$\mathcal{S}=(1+\mathrm{i}\mathcal{R})(1-\mathrm{i}\mathcal{R})^{-1}$.
The matrix $\mathcal{S}$ is unitary, but 
(again unlike conventional theory)
asymmetrical.
The 
basis of outgoing waves is obtained from the
real basis
 by transformation
\beq
   g_{n''s'';\,\kappa}^\mathrm{out}(z) = 2\mathrm{i}
   \sum_{n's'}\big[(1+\mathrm{i}\mathcal{R})^{-1}\big]_{ns;\,n's'}
   \,g_{n''s'';\,\kappa'}^\mathrm{real}(z).
\label{g_as_out}
\eeq
Here, pairs $(n,s)$ and $(n',s')$, being respective
parts of the composite
quantum numbers $\kappa$ and $\kappa'$, run over
open channels, but $(n'',s'')$ run over all 
(open and closed) channels.
From the unitarity of the 
scattering matrix it follows that 
the wave functions that satisfy the asymptotic condition 
given by \req{asout} should be multiplied by 
a common factor $(2L_z)^{-1/2}$,
to ensure the normalization expressed by \req{L_z}. 
As the initial state of 
a transition, one should use the basis of incoming waves,
$g_{n''s'';\,\kappa}^\mathrm{in}(z)=
[g_{n''s'';\,\kappa}^\mathrm{out}(z)]^\ast$.

After the
ortho-normalized outgoing waves have been constructed
for each $z$-parity, with symmetric
and antisymmetric longitudinal coefficients
$g_{n''s'';\,ns\vec{K}E\pm}^{\mathrm{out}}(z)=\pm
g_{n''s'';\,ns\vec{K}E\pm}^{\mathrm{out}}(-z)$ 
in expansion (\ref{expan}),
solutions for electron waves propagating
at $z\to\pm\infty$
in a definite open channel $(n,s)$
with a definite momentum 
$\hbar k=(\mathrm{sign}\,z)\,\hbar k_{ns}$ 
are given by the expansion in \req{expan}
with coefficients
\beq
g_{n''s'';\,nsk}(z)=
   (g_{n''s'';\,nsE+}^\mathrm{out}(z)\pm
 g_{n''s'';\,nsE-}^\mathrm{out}(z)/{\sqrt{2}},
\label{k-basis}
\eeq
where the sign $+$ or $-$ represents electron
escape in the positive or negative
$z$ direction, respectively,
and we have suppressed $\vec{K}$ in the subscripts.
Waves incoming from $z\to\pm\infty$
with a definite momentum 
are given by the complex conjugate of \req{k-basis}.

\subsection{Adiabatic approximation}

In early works on the H atom in strong magnetic fields, a so-called
adiabatic approximation was widely used (e.g.,
\citealt{GorDzyal,CanutoVentura} and references
therein), which neglects all terms
but one in \req{expan}, i.e.,
\beq
  \psi_\kappa(\vec{r})=\Phi_{ns}(\vec{r}_\perp)\,g_\kappa(z).
\label{adiabatic}
\eeq
This approximation reduces the system (\ref{Hsystem})
to the
single equation with $(n'',s'')=(n,s)$ and zero r.h.s.

The accuracy of the adiabatic approximation for bound states can
be assessed by comparing $|E^\|_\kappa|$ with the distance
between the neighboring Landau levels that are coupled by the
r.h.s.\ of \req{Hsystem}. For an atom \emph{at rest} ($K=0$), all
the channel-coupling terms become zero for $s''\neq s$. In this
case, the relevant Landau level distance is $\hbar\omce$, while
the longitudinal energies of the states with $\nu=0$ (``tightly
bound states'') are $|E^\|_{ns0}(0)|\sim 0.1$--$0.3$ keV at
$B\sim10^{11}$--$10^{14}$~G , so that the adiabatic approximation
is accurate to within a few percent or better. It becomes still
better for the ``hydrogenlike states'' with $\nu>0$, which have
$|E^\|_\kappa|\lesssim 0.01$ keV.

From comparison of $|E^\|_{ns\nu}(0)|$ with $\hbar\omcp$,
one can conclude that the adiabatic approximation is
generally inapplicable to a \emph{moving} atom.
However, the accuracy remains good for sufficiently slow atoms,
that is when either $\kp\ll\kc$ or (provided that $\eta=1$)
$\kp\gg\kc$ \citep{P94,P98}.
Otherwise, since off-diagonal
effective potentials $V_{n''s'',n's'}(r_B,z)$ in \req{Hsystem}
decrease at $|z|\to\infty$ more rapidly than
diagonal ones, this approximation
accurately reproduces wave functions tails at large $|z|$,
provided that $\eta=1$.

For continuum states,
the reactance and scattering matrices are diagonal
 in the adiabatic approximation,
 with a separate scattering coefficient
$S_{ns}=S_{ns;\,ns}$ for every open channel.

\subsection{Born approximation}
\label{sect:Born}

In the Born approximation, the
potential $V$ in a Hamiltonian $H=H_0+V$, which acts on particles
in the continuum states, is treated as a small perturbation. We
define $\psi^{(0)}$ to be the nonperturbed function, which satisfies the
equation $H_0\psi^{(0)}=E\psi^{(0)}$. Then from the Schr\"odinger
equation $H\psi=E\psi$, one obtains the continuum wave function in
the first Born approximation in the form
$\psi=\psi^{(0)}+\psi^{(1)}$, where $\psi^{(1)}$ is determined by
the equation
$
(H_0-E)\psi^{(1)}=-V\psi^{(0)}.
$

Since we consider the continuum states  corresponding to definite
Landau numbers $(n,N)$ at $|z|\to\infty$ (Sect.\,\ref{sect:WFC}), 
the zero-order wave
function $\psi^{(0)}$ is given by the adiabatic approximation
with $g_\kappa(z)$ replaced by plane
waves.

\subsubsection{Two forms of solution}

We now consider continuum states. 
We first choose the nonperturbed wave function
in the representation where the $z$-projections of
angular momentum operators,
$\ell_z$, have definite values $-\hbar s_\mathrm{e}$ for
the electron
and $\hbar s_\mathrm{p}$ for the proton. Then
\beq
  \psi^{(0)}_{n,s_\mathrm{e},k_{z\mathrm{e}},N,s_\mathrm{p},k_{z\mathrm{p}}}(\vec{r}_\mathrm{e},\vec{r}_\mathrm{p}) =
   \Psi_{n,s_\mathrm{e},k_{z\mathrm{e}}}(\vec{r}_\mathrm{e})\Psi_{N,s_\mathrm{p},k_{z\mathrm{p}}}(\vec{r}_\mathrm{p}),
\label{psi0ep}
\eeq
and $\psi^{(1)}$ is governed by the equation
\beq
   \big( H^{(1)}_{e} + H^{(1)}_{p}-E \big) \,
   \psi^{(1)}(\vec{r}_\mathrm{e},\vec{r}_\mathrm{p}) =
   ({e^2}/{r_{\mathrm{ep}}})\,\,\psi^{(0)}(\vec{r}_\mathrm{e},\vec{r}_\mathrm{p}).
\label{Born-H}
\eeq
Using an expansion of $\psi^{(1)}$
over the complete set of 
$\psi^{(0)}_{n',s'_\mathrm{e},k'_{z\mathrm{e}},N',s'_\mathrm{p},k'_{z\mathrm{p}}}$,
we obtain in the standard way 
\bea
   \psi^{(1)}(\vec{r}_\mathrm{e},\vec{r}_\mathrm{p}) 
   &=&
\frac{\mathrm{e}^{\mathrm{i}k_z(z_\mathrm{e}-z_\mathrm{p})}}{L_z^2}
 \!\!\sum_{n',s',N',s'',k'_z}\!\!
  W^{n's'N's''}_{n,s_\mathrm{e},N,s_\mathrm{p}}(k_z-k'_{z})
\nonumber\\&&\,\,\,\times\,
  \frac{\Phi_{n's'}(\vec{r}_{\perp\mathrm{e}})
       \,\Phi^\ast_{N's''}(\vec{r}_{\perp\mathrm{p}})
     }{E^\perp_{n'}+E^\perp_{N'} + (\hbar k'_{z})^2/2\mred - E
       } ,
\label{psi1ep}
\eea
where 
$k_z=(k_{z\mathrm{e}}\mpr-k_{z\mathrm{p}}\mel)/\mH=k_{z\mathrm{e}}=-k_{z\mathrm{p}}$ (since $P_z=0$)
and
\bea&&
   W^{n's'N's''}_{n,s_\mathrm{e},N,s_\mathrm{p}}(k) = e^2
   \int_{-\infty}^\infty\!\!\!\!\dd z\,
     \mathrm{e}^{\mathrm{i}kz}
   \int_{\mathbb{R}^2\otimes\mathbb{R}^2}
   \dd\vec{r}'_\perp\,\dd\vec{r}_\perp\,
\nonumber\\&&\quad\times
   \frac{
    \Phi^\ast_{n's'}(\vec{r}'_\perp)
    \,\Phi_{N's''}(\vec{r}_\perp)
    \,\Phi_{ns_\mathrm{e}}(\vec{r}'_\perp)
    \,\Phi^\ast_{Ns_\mathrm{p}}(\vec{r}_\perp)
}{\big(|\vec{r}'_\perp-\vec{r}_\perp|^2+z^2\big)^{1/2}}.
\qquad
\label{Wep}
\eea
In the limit
$L_z\to\infty$, we replace $\sum_{k_z}$ by
$(L_z/2\pi)\int_{-\infty}^\infty \dd k_z$. 

\citet{PC03} obtained a simpler solution, based on the
representation of quantum states with definite $\vec{K}$. In this
case,  there are no separate quantum numbers $s_\mathrm{e}$ and $s_\mathrm{p}$. 
After applying the
 transformation in \req{PsiH}, $\psi^{(0)}$
is given by \req{adiabatic} with
$g_{\kappa}(z)=\mathrm{e}^{\pm\mathrm{i}k_zz}/\sqrt{L_z}$. 
Using Fourier transform
\beq
   \tilde\psi(\vec{r}_\perp,k) = \frac{1}{\sqrt{L_z}}
   \int_{-L_z/2}^{L_z/2}
   \mathrm{e}^{-\mathrm{i}kz}\,\psi(\vec{r})\,\dd z,
\eeq
we obtain from \req{Hsystem} in the
first Born approximation
\beq
   \tilde\psi_\kappa(\vec{r}_\perp,k)
     =\tilde\psi_\kappa^{(0)}(\vec{r}_\perp,k)
     +\tilde\psi_\kappa^{(1)}(\vec{r}_\perp,k)
\eeq
with
\bea
\tilde\psi_\kappa^{(0)}(\vec{r}_\perp,k) 
&=& \frac{\sin((k-k_z)L_z/2)}{(k-k_z)L_z/2}\,\Phi_{ns}(\vec{r}_\perp)
\nonumber\\
&\simeq&
({2\pi}/{L_z})\,\delta(k-k_z)\,
\Phi_{ns}(\vec{r}_\perp)
\mbox{~at~}L_z\to\infty,
\\
   \tilde\psi_\kappa^{(1)}(\vec{r}_\perp,k)
    &=& \sum_{n's'} \tilde{g}^{(1)}_{n's';\,\kappa}(k)\,
   \Phi_{n's'}(\vec{r}_\perp),
\label{psi1}
\\
   \tilde{g}^{(1)}_{n's';\,\kappa}(k)
   &=& \frac{- L_z^{-1}\,\tilde{V}_{ns}^{n's'}(r_c, k-k_z)
   }{\displaystyle
   \hbar^2 (k^2-k_z^2)/(2\mred) + E_{n's'}^\perp
   - E_{ns}^\perp
   }\,,
\label{system-Fourier}
\\
   \tilde{V}_{ns}^{n's'}(r_c, k) &=&
   \int_{-\infty}^\infty \mathrm{e}^{-\mathrm{i} kz}
   \,V_{ns,n's'}(r_c,z)\,\dd z,
\label{V-Fourier}
\eea
where $\tilde{V}_{ns}^{n's'}$
can be presented
as a single integral of a combination of elementary
functions (Appendix B of \citealt{PC03}).

\subsubsection{Approximation of infinite proton mass}
\label{sect:infimass}

The neglect of the proton motion is equivalent to the
assumption that $\mpr\to\infty$. In this approximation,
$\psi^{(0)}$ depends only on $\vec{r}_\mathrm{e}$ in \req{psi0ep}
without $\Psi_{N,s_\mathrm{p},k_{z\mathrm{p}}}(\vec{r}_\mathrm{p})$
on the r.h.s.
Then \req{psi1ep} simplifies to
\bea&&
   \psi^{(1)}(\vec{r}_\mathrm{e}) 
   =
   \sum_{n',s',k'_z}
  \frac{\mathrm{e}^{\mathrm{i}k_z z_\mathrm{e}}
     }{L_z}
    \,\frac{\Phi_{n's'}(\vec{r}_{\perp\mathrm{e}})
 }{E^\perp_{n'}
      + (\hbar k'_{z})^2/2\mel - E}
   \!\int_{-L_z}^{L_z}\!\dd z'\,
     \frac{\mathrm{e}^{\mathrm{i}(k_z-k'_{z})z'}
       }{L_z}
\nonumber\\&&\quad\times
   \int_{\mathbb{R}^2}
   \dd\vec{r}'_\perp
    \Phi^\ast_{n's'}(\vec{r}'_\perp)
    \,\frac{e^2}{\sqrt{|\vec{r}'|^2+(z')^2}}
    \, \Phi_{ns_\mathrm{e}}(\vec{r}'_\perp).
\label{psi1e}
\eea
Taking into account the definition in \req{Veff}, we see that this
solution is identical to the solution provided by 
Eqs.~(\ref{psi1})--(\ref{V-Fourier}) in the particular case where
$r_c=0$, after the obvious replacement of $E^\perp_{ns}$ by
$E^\perp_n$ and $\mred$ by $\mel$. The zero value of $r_c$
naturally reflects the condition $v_\mathrm{drift}=0$.

\section{Electron-proton photoabsorption}
\label{sect:ff}

\subsection{General expressions}
\label{sect:gen}

The general nonrelativistic formula for the differential cross-section
of absorption of radiation by a quantum-mechanical system
 is (e.g., \citealp{Armstrong})
\beq
   \dd\sigma=\frac{4\pi^2}{\omega c}
   \left|\langle f| \vec{e}\cdot\vec{j}_\mathrm{eff} | i \rangle
   \right|^2\,\delta(E_f-E_i-\hbar\omega)\,\dd\nu_f,
\label{dsigma}
\eeq
where $|i\rangle$ and $|f\rangle$ are the initial and final states of 
the system, $\dd\nu_f$ is the density of final states,
$\hbar\omega$ is the photon energy,
$\vec{e}$ is the polarization vector, 
$\vec{j}_\mathrm{eff}=
\mathrm{e}^{\mathrm{i}\vec{q}\cdot\vec{r}} \vec{j}$,
$\vec{j}$ is the electric current operator, and $\vec{q}$
is the photon wave number.
In our case,
\beq
   \vec{j}_\mathrm{eff}=
   e\,(\mathrm{e}^{\mathrm{i}\vec{q}\cdot\vec{r}_\mathrm{e}}
   \dot{\vec{r}_\mathrm{e}} -
   \mathrm{e}^{\mathrm{i}\vec{q}\cdot\vec{r}_\mathrm{p}}
   \dot{\vec{r}_\mathrm{p}}),
\label{jeff-gen}
\eeq
where the velocity operators $\dot{\vec{r}}_{e}$ and
$\dot{\vec{r}}_{p}$ are given by \req{r-dot}.

Equation (\ref{jeff-gen}) 
does not yet include either the photon interaction with 
electron and proton magnetic moments $\hat{\vec{\mu}}_\mathrm{e}$
or $\hat{\vec{\mu}}_\mathrm{p}$. 
For transitions without spin-flip, the latter interaction
can be taken into account by
adding to the $\vec{e}\cdot\vec{j}_\mathrm{eff}$ operator the term
$
   \hat{j}_\mathrm{spin} =
     -\mathrm{i}
   (\vec{q}\times\vec{e})\cdot(\hat{\vec{\mu}}_\mathrm{e}
   + \hat{\vec{\mu}}_\mathrm{p})
$
(cf.\ \citealt{KVH}),
whereas operators
$(\vec{q}\times\vec{e})\times\hat{\vec{\mu}}_{e,p}$
are responsible for spin-flip transitions 
(cf.\ \citealt{Wunner-ea83}).

We consider the representation where $s_\mathrm{e}$ and $s_\mathrm{p}$ are definite in
the initial and final states. For an initial state with fixed
$n_i$, $s_{e,i}$, $N_i$, $s_{p,i}$, and $k_z=k_i$ in
\req{psi1ep}, and for a final state with either a fixed
$z$-parity or a fixed sign of $k_z=k_f$,  we
have in \req{dsigma} $\dd\nu_f=(L_z/2\pi)\,(\mred/\hbar^2
|k_f|)\,\dd E_f$. Therefore, the cross-section of
photoabsorption for a pure initial
quantum state $|i\rangle$ is
\beq
   \sigma_i(\omega) =
   \hspace*{-2em}
   \sum_{n_f,s_{e,f},N_f,s_{p,f},\pm\rule{0pt}{2ex}}
    \hspace*{-1em} 
   \frac{2\pi L_z \mred}{\hbar^2 |k_f| \omega\,c}\,
   \left|\langle f |
    \mathcal{J}
    | i \rangle
   \right|^2 ,
\label{partial-ep}
\eeq
where
$\mathcal{J}\equiv\vec{e}\cdot\vec{j}_\mathrm{eff} + \hat{j}_\mathrm{spin}$,
\beq
   E_{n_f s_f}^\perp + \frac{\hbar^2 k_f^2}{2\mred}
   = E_{n_i s_i}^\perp + \frac{\hbar^2 k_i^2}{2\mred} + \hbar\omega ,
\label{en-conserv}
\eeq
the sum is performed over those $n_f$ and $N_f$ which are
permitted by \req{en-conserv},
and ``$\pm$'' means the sum over the $z$-parity of the final state
(in the case where the parity is definite)
or the sum over the signs of $k_f$ 
(in the case where $k_z$ in the final state is definite).

In the alternative representation with definite
cartesian components of pseudomomentum
$\vec{K}$, using the transformation
in \req{PsiH}, 
one can express the cross-section in terms of the interaction
matrix element between the initial and final
\emph{internal} states of the electron-proton system
\citep{BP94}.
The result has the same form as \req{dsigma}, but now
$\dd\nu_f$ is the density of final states at fixed
$\vec{K}_f=\vec{K}_i+\hbar\vec{q}$, initial and final states
are described by wave functions $\psi_{\vec{K}}$,
and the effective current operator in the
conventional representation with
$\eta=0$ ($\vec{r}=\vec{r}_{\mathrm{ep}}$)
is given by
\bea
      \vec{j}_\mathrm{eff} &=& e\,\exp\left(\mathrm{i}
   \frac{\mpr}{\mH}\vec{q}\cdot\vec{r} \right)
 \left(\frac{\vec{\pi}}{\mel} 
   + \frac{\vec{K}}{\hbar\mH} 
   + \frac{\hbar\vec{q}}{2\mel} \right)
\nonumber\\&&
   + \, e\,\exp\left(-\mathrm{i}
   \frac{\mel}{\mH}\vec{q}\cdot\vec{r} \right)
 \left(\frac{\vec{\Pi}}{\mpr} 
   - \frac{\vec{K}}{\hbar\mH}
   - \frac{\hbar\vec{q}}{2\mpr} \right)
   ,\quad
\label{hat-M}
\eea
where operator $\vec{\pi}$ is defined by \req{pi} and
$
   \vec{\Pi} = \vec{p}-({e}/{2c})\,\vec{B}\times\vec{r}.
$
The transverse cyclic components of operators
$\vec{\pi}$ and $\vec{\Pi}$
act on the Landau states as
\bea
   \pi_{\pm1} |n,s\rangle_\perp &=& \mp \frac{\mathrm{i}\hbar}{\am}
   \sqrt{n^\pm}\,|n\pm1,s\mp1\rangle_\perp,
\label{pi+-}
\qquad\\
   \Pi_{\pm1} |n,s\rangle_\perp &=& \mp \frac{\mathrm{i}\hbar}{\am}
   \sqrt{N^\mp}\,|n,s\mp1\rangle_\perp,
\label{Pi+-}
\qquad
\eea
where
$n^\pm\equiv n+\frac12\pm\frac12$
and $N^\mp\equiv
N+\frac12\mp\frac12=n^\mp+s$.

Changes in $\vec{r}_A$ and $\vec{r}_B$ induce 
transformations of operator $\vec{j}_\mathrm{eff}$,
studied by \citet{BP94}.
In the particular case where for both initial and final states,
the representation with $\eta=1$ ($\vec{r}_B=\vec{r}_c$,
$\vec{r}=\vec{r}_{\mathrm{ep}}+\vec{r}_c$)
is used, their result reads
\bea
      \frac{\vec{j}_\mathrm{eff}}{e} &=& 
\exp\left(\mathrm{i}\vec{q}\cdot\frac{\vec{r}_\perp+\vec{r}_c}{2}
    + \frac{\mpr}{\mH}\,\mathrm{i} q_z z \right)
   \,
   \left(\frac{\vec{\pi}}{\mel}+\frac{\hbar\vec{q}}{2\mel}\right)
\nonumber\\&&
   +
\exp\left(-\mathrm{i}\vec{q}\cdot\frac{\vec{r}_\perp+\vec{r}_c}{2}
    - \frac{\mel}{\mH}\,\mathrm{i} q_z z\right)
\,
   \left(\frac{\vec{\Pi}}{\mpr}-\frac{\hbar\vec{q}}{2\mpr}
   \right).
   \quad
\label{m_shift}
\eea
In this representation, instead of \req{partial-ep}, we have
\beq
   \sigma_i(\omega) =
   \sum_{n_f,s_f,\pm} 
   \frac{2\pi L_z \mred}{\hbar^2 |k_f| \omega\,c}\,
   \left|\langle f |
    \mathcal{J}
    | i \rangle
   \right|^2 ,
\label{partial}
\eeq
where the sum is performed over those $n_f$ and $s_f$ that are
permitted by \req{en-conserv}.
For the solution described in Sect.\,\ref{sect:exact},
the matrix element in \req{partial} becomes
\bea
   \langle f |
    \mathcal{J}
    | i \rangle &=&
   \sum_{n',s',n'',s''} 
    \int_{-L_z}^{L_z} \!\!
   \big[g_{n''s'';\,\kappa_f}^{\mathrm{out}}(z)
   \big]^\ast
\nonumber\\&&\times
   \langle n''s''\,|\,\mathcal{J}\,|\,n's'\rangle_\perp\,
   g_{n's';\,\kappa_i}^{\mathrm{in}}(z)\,\dd z.
\label{J-sum}
\eea
Using Eqs.~(\ref{pi+-}) and (\ref{Pi+-}), we can express
the transverse matrix elements 
$\langle n''s''\,|\,\mathcal{J}\,|\,n's'\rangle_\perp$
in terms of Laguerre 
functions.
Hence, \req{J-sum} presents
a sum of overlap integrals over $z$.
For instance, Eqs.~(A7)--(A12) of \citet{PP97}
provide an explicit expression in terms of
this overlap integrals
for the matrix elements $\langle f \,|\,\hat{M}\,|\, i \rangle$
of the operator
$\hat{M}=(\hbar/e^3)\,\mathcal{J}$
in the approximation
where small terms $\sim O((\mel/\mpr)\,q)$ are neglected, but
separate terms $\sim O(\mel/\mpr)$ and $\sim O(q)$ are retained.

\subsection{Dipole and Born approximations}

Hereafter, we use the dipole approximation ($q\to0$).
Then  
$\hat{j}_\mathrm{spin}$ vanishes, and the total effective
current in \req{jeff-gen} reduces to
\beq
   \vec{j}_\mathrm{eff} = e(\dot{\vec{r}}_\mathrm{e}-\dot{\vec{r}}_\mathrm{p}),
\eeq
while
the transformed effective current in \req{m_shift} becomes
\beq
 \vec{j}_\mathrm{eff} = e \left({\vec{\pi}}/{\mel}
 + {\vec{\Pi}}/{\mpr} \right).
\label{jeff}
\eeq
By substituting Eqs.~(\ref{pi+-}) and (\ref{Pi+-}),
the sum in \req{J-sum} reduces to
\beq
   \langle f \,|\,
    \mathcal{J}
    \,|\, i \rangle = \sum_{\alpha=-1}^{+1} \, \sum_{n's'} 
e_{-\alpha} \,\bar{j}^{(\alpha)}_{n's'}(\kappa_i,\kappa_f),
\eeq
where
\bea&&
  \bar{j}^{(+1)}_{n's'} = -\frac{\mathrm{i}\hbar e}{\am}
  \bigg(\frac{\sqrt{n'+1}}{\mel}\mathcal{I}^{n'+1,N'}_{n'N'}
  \!
   + 
   \frac{\sqrt{N'}}{\mpr}\mathcal{I}^{n',N'-1}_{n'N'}\bigg),
\qquad\\&&
  \bar{j}^{(-1)}_{n's'} = \frac{\mathrm{i}\hbar e}{\am}
  \bigg(\frac{\sqrt{n'}}{\mel}\mathcal{I}^{n'-1,N'}_{n'N'} + 
   \frac{\sqrt{N'+1}}{\mpr}\mathcal{I}^{n',N'+1}_{n'N'}\bigg),
\\&&
   \mathcal{I}^{n'',N''}_{n',N'} = 
   \int_{-L_z}^{L_z}
    \left[g_{n'',s'';\,\kappa_f}^{\mathrm{out}}(z)\right]^\ast
   g_{n's';\,\kappa_i}^{\mathrm{in}}(z)\,\dd z,
\\&&
  \bar{j}^{(0)}_{n's'} = -{\mathrm{i}\hbar e}
   \int_{-L_z}^{L_z}
    \left[g_{n's';\,\kappa_f}^{\mathrm{out}}(z)\right]^\ast
    \frac{\dd}{\dd z}
   g_{n's';\,\kappa_i}^{\mathrm{in}}(z)\,\dd z,
\eea
and $N\equiv n+s$ is the proton Landau number.
In
the first Born approximation (Sect.\,\ref{sect:Born}),
\beq
   \langle f\,|\,\mathcal{J}\,|\,i  \rangle
   \approx
         \langle\psi^{(1)}_f\,|\,
   \vec{e}\cdot\vec{j}_\mathrm{eff}
   \,|\,\psi^{(0)}_i\rangle
+
   \langle\psi^{(0)}_f\,|\,
   \vec{e}\cdot\vec{j}_\mathrm{eff}
   \,|\,\psi^{(1)}_i\rangle.
\label{Born}
\eeq
In the representation where $s_\mathrm{e}$ and $s_\mathrm{p}$ are definite,
using \req{psi0ep} for $\psi^{(0)}$
and \req{psi1ep} for $\psi^{(1)}$, and
taking into account the relations in \req{pi1+-},
we can derive the explicit expression for
the matrix element in \req{partial-ep} of
\beq
   \langle f\,|\,
    \vec{e}\cdot\left(\dot{\vec{r}}_\mathrm{e}-\dot{\vec{r}}_\mathrm{p}\right)
    \,|\,i  \rangle
   = \sum_{\alpha=-1}^{+1}
e_{-\alpha} \,
\bar{j}^{(\alpha)}_{n_f,s_{e,f},N_f,s_{p,f};\,
n_i,s_{e,i},N_i,s_{p,i}},
\eeq
where
\bea
   \bar{j}^{(0)}
   &=& -\frac{e}{L_z\,\mred\,\omega}\,\Delta k\,
   W^{n_f,s_{e,f},N_f,s_{p,f}}_{n_i,s_{e,i},N_i,s_{p,i}}(\Delta k) ,
\\
   \bar{j}^{(\pm1)}
   &=&    \pm \frac{\mathrm{i} e}{ L_z \am}\,\bigg\{
   \frac{1}{\mel\,(\omega \pm \omce)}\,
   \bigg(\sqrt{n_f^\pm}\,
   W^{n_f\pm1,s_{e,f}\mp1,N_f,s_{p,f}}_{n_i,s_{e,i},N_i,s_{p,i}}(\Delta k)
\nonumber\\&&\quad
   - \sqrt{n_i^\mp}\,
   W^{n_f,s_{e,f},N_f,s_{p,f}}_{n_i\mp1,s_{e,i}\pm1,N_i,s_{p,i}}(\Delta k)
   \bigg)
\nonumber\\&& +
   \frac{1}{\mpr\,(\omega \mp \omcp)}\,
   \bigg(\sqrt{N_f^\mp}\,
   W^{n_f,s_{e,f},N_f\mp1,s_{p,f}\pm1}_{n_i,s_{e,i},N_i,s_{p,i}}(\Delta k)
\nonumber\\&&\quad
   - \sqrt{N_i^\pm}\,
W^{n_f,s_{e,f},N_f,s_{p,f}}_{n_i,s_{e,i},N_i\pm1,s_{p,i}\mp1}(\Delta k)
   \bigg)\bigg\},
\eea
and $\Delta k\equiv k_f-k_i$.

In the representation where 
cartesian components of $\vec{K}$ have definite values,
using Eqs.~(\ref{psi1})--(\ref{V-Fourier}),
(\ref{pi+-}), and (\ref{Pi+-}),
one can derive the matrix element in \req{partial}
in the form
\beq
   \langle f\,|\,\mathcal{J}\,|\,i\rangle = \sum_{\alpha=-1}^{+1}
e_{-\alpha} \,
\bar{j}^{(\alpha)}_{n_f,s_f;\,
n_i,s_i},
\label{alphasum}
\eeq
where
\bea
   \bar{j}^{(0)}
   &=& -\frac{ e }{L_z\,\mu\,\omega}\,\Delta k\,
   \tilde{V}_{n_i s_i}^{n_f s_f}   (r_c,\Delta k) ,
\\
   \bar{j}^{(\pm1)}
   &=& \!
   \pm\frac{\mathrm{i} e}{ L_z \am} \left(
   \frac{ \sqrt{n_f^\mp}\,\tilde{V}_{n_i s_i}^{n_f \mp1, s_f\pm1}
   (r_c,\Delta k)
   - \!\sqrt{n_i^\pm}\,\tilde{V}_{n_i\pm1, s_i\mp1}^{n_f s_f}
   (r_c,\Delta k)
   }{ \mel\,(\omega \pm \omce)}
\right.\nonumber\\&&\quad\left. +
   \frac{ \sqrt{N_f^\pm}\,\tilde{V}_{n_i s_i}^{n_f, s_f\pm1}
   (r_c,\Delta k)
   - \sqrt{N_i^\mp}\,\tilde{V}_{n_i, s_i\mp1}^{n_f s_f}
   (r_c,\Delta k)
   }{ \mpr\,(\omega \mp \omcp)} \right).
\label{j+-}
\eea
Substituting Eqs.~(\ref{alphasum})--(\ref{j+-}) into
\req{partial},
assuming Maxwell distribution of $k_i$,
 and taking the average over the initial states,
 we obtain \citep{PC03,PL07}
\beq
   \sigma(\omega) = \sum_{\alpha=-1}^{+1}|e_{\alpha}|^2
   \sum_{n,N}f^e_n f^p_N \sum_{n',N'}
   \,\sigma^{(\alpha)}_{n,N;\,n',N'}(\omega),
\label{sigmasum}
\eeq
where $f^e_n$ and $f^p_N$ are the electron and proton
number fractions at the Landau levels $n$ and $N$,
\beq
   \sigma^{(\alpha)}_{n,N;\,n',N'}(\omega) =
          \frac{4\pi e^2
          }{ 
     \mel c} \,
   \frac{\omega^2\,\nu^{(\alpha)}_{n,N;\,n',N'}(\omega)
          }{
          (\omega+\alpha\omce)^2 \, (\omega-\alpha\omcp)^2
             }
\label{sigmaNN}
\eeq
is the partial cross-section for transitions between the
specified electron and proton Landau levels
for polarization $\alpha$,
\beq
   \nu^{(\alpha)}_{n,N;\,n',N'}(\omega) = 
        \frac{4}{3}\,\sqrt{\frac{2\pi}{\mel T}}\,
          \frac{n_\mathrm{e}\, e^4}{\hbar \omega}\,
 \Lambda_{n,N;\,n',N'}^{(\alpha)}(\beta_\ast,\omega/\omega_\ast)
\label{Gamma-nNn'N'}
\eeq
is the effective partial collision frequency,
\bea&&\hspace*{-1em}
   \Lambda_{n,N;\,n',N'}^{(\alpha)}
       (\beta_\ast,\omega/\omega_\ast)
   = \frac32 \int_0^\infty \frac{\dd u}{u'}\,\mathrm{e}^{-\beta_\ast u^2/2}
   \,\theta({u'}^2)\,
\nonumber\\&&\times\,
    \left((u'+ u)^{2|\alpha|}\, w_{n,N;\,n',N'}^{(\alpha)}(u_+)
    + (u'- u)^{2|\alpha|}\, w_{n,N;\,n',N'}^{(\alpha)}(u_-)
    \right)
\label{Lambda0}
\eea
is a partial Coulomb logarithm,
and
\beq
       w_{n,N;\,n',N'}^{(\alpha)}(u_\pm)
    = \frac12 \int_0^\infty \frac{t^{\,|\alpha|}\,\dd t}{(t+u_\pm^2/2)^2}
   \,I_{n',n}^2(t)\,
   I_{N',N}^2(t) ,
\label{wN1N2a}
\eeq
where 
$\beta_\ast = \hbar\omega_\ast/T = \hbar eB/(m_\ast c T)$,
$u_\pm = |u\pm u'|$, 
$\theta({u'}^2)$ is the Heaviside step function,
and
\beq
   {u'}^2 = u^2 + \frac{2m_\ast}{\mpr}\,(N-N')
    + \frac{2m_\ast}{\mel}\,(n-n')
    + \frac{2\omega}{\omega_\ast}.
\label{u'ff}
\eeq
Since $\Lambda^{(+1)}=\Lambda^{(-1)}$,
two different Coulomb logarithms
$\Lambda^{(0)}$ and $\Lambda^{(\pm1)}$ 
describe all three basic polarizations.

Terms that are proportional to
$e_{\alpha\phantom{'}}\! e_{\alpha'}^\ast$ with $\alpha\neq\alpha'$
are absent in \req{sigmasum}, because, for every pair of pure
quantum states $|i\rangle$ and $|f\rangle$, 
only one of the three basic
polarizations provides a non-zero transition matrix element
in the dipole approximation.

\citet{PL07} mentioned that Debye screening
might be taken into account by using
$u_\pm = [(u\pm u')^2 + (\am k_\mathrm{D})^2]^{1/2}$ 
as the arguments of
$w_{n,N;\,n',N'}^{(\alpha)}$ 
in \req{Lambda0}, 
$k_\mathrm{D}$ being the inverse screening length. However,
\citet{sawyer07}, following \citet{Bekefi}, showed that
scattering off a Debye potential 
is not a valid description of the screening
correction for photoabsorption; instead, the integrand in
\req{wN1N2a} should be multiplied by
$
   (t+u_\pm^2 + \am^2 k_\mathrm{De}^2/2)
   /
   (t+u_\pm^2 + \am^2 k_\mathrm{D}^2),
$
where $k_\mathrm{De}^2$ is the electron contribution
to the squared Debye wave number $k_\mathrm{D}^2$.

\subsection{Damping factor}
\label{sect:damp}

Equation (\ref{sigmaNN}) gives divergent results
at $\omega\to\omce$ for $\alpha=-1$ and at $\omega\to\omcp$ for
$\alpha=+1$, because it ignores damping effects due to the finite
lifetimes of the initial and final states of the transition.
A conventional
way of including these effects consists of adding
a damping factor to the denominator in \req{sigmaNN},
which results in Lorentz profiles \citep[e.g.,][]{Armstrong}.
The damping factor can be
traced back to the accurate treatment of the complex dielectric
tensor of the classical magnetized plasma \citep{Ginzburg}.
This treatment allows one to express the 
complex dielectric tensor
in terms of the effective collision
frequencies related to 
different types of collisions in the plasma.
Imaginary parts of the refraction indexes,
calculated from the complex dielectric tensor,
provide complicated expressions for the free-free
photoabsorption cross sections $\sigma_\alpha^\mathrm{ff}$
for the basic polarizations $\alpha=0,\pm1$.
Based on the assumption that the effective collision
frequencies are small compared to $\omega$,
the latter expressions
greatly simplify and reduce to \citep{PC03}
\beq
     \sigma_\alpha^\mathrm{ff}(\omega) = 
             \frac{4\pi e^2
          }{ 
     \mel c} \,
   \frac{\omega^2\,\nu_\alpha^\mathrm{ff}(\omega)
          }{
          (\omega+\alpha\omce)^2 \, (\omega-\alpha\omcp)^2
             +\omega^2 \, \tilde\nu_\alpha^2(\omega)},
\label{sigma-fit}
\eeq
where
\beq
     \tilde\nu_\alpha =
     \left(1+\alpha\,\frac{\omce}{\omega}\right)
     \nu_\mathrm{p}
     + \left(1-\alpha\,\frac{\omcp}{\omega}\right)
     \nu_\mathrm{e} + \nu_\alpha^\mathrm{ff},
\label{damp}
\eeq
$\nu_\mathrm{e}$ and $\nu_\mathrm{p}$ being the
effective damping factors for protons and electrons,
respectively, not related to the electron-proton collisions.
In general, 
$\nu_\mathrm{e}$ and $\nu_\mathrm{p}$ may also depend on $\alpha$ and $\omega$.
\citet{Ginzburg} considers $\nu_\mathrm{e}$ and $\nu_\mathrm{p}$
for collisions
of electrons and protons with molecules, whereas \citet{PC03}
take into account damping factors due to both the
scattering of light by free electrons and protons
and proton-proton collisions.
The derivation of \req{damp}
from the complex dielectric tensor of the plasma
assumes that $\nu_{e}\ll\omce$, $\tilde\nu_\alpha\ll\omce$,
and $\nu_\mathrm{p}\ll\omcp$.

Although the general expressions given in Eqs.~(\ref{sigma-fit}),
(\ref{damp}) can be established
in frames of the classical theory, 
accurate values of the effective frequencies are provided
by quantum mechanics. In our case, 
\beq
   \nu_{\alpha}^{\mathrm{ff}}(\omega) =
    \sum_{n,N} f^e_n f^p_N \sum_{n',N'}
\nu_{n,N;\,n',N'}^{(\alpha)}(\omega) 
=
        \frac{4}{3}\,\sqrt{\frac{2\pi}{\mel T}}\,
          \frac{n_\mathrm{e}\, e^4}{\hbar \omega}
 \Lambda_{\|,\perp}^{\mathrm{ff}},
\label{nu-ff}
\eeq
where $\nu_{n,N;\,n',N'}^{(\alpha)}(\omega)$ is provided by
Eqs.~(\ref{Gamma-nNn'N'})--(\ref{u'ff}). In the second
equality,
$\Lambda_\|^{\mathrm{ff}}$ and $\Lambda_\perp^{\mathrm{ff}}$
are, by definition,
Coulomb logarithms for $\alpha=0$ and $\alpha=\pm1$. 
Parallel and transverse  Gaunt factors (e.g.,
\citealt{Mesz}) equal
$(\sqrt{3}/\pi)\,\Lambda_{\|}^{\mathrm{ff}}$ and
$(\sqrt{3}/\pi)\,\Lambda_{\perp}^{\mathrm{ff}}$, respectively.

Since different quantum transitions contribute to the cyclotron
resonance at the same frequency ($\omcp$ or $\omce$, depending on
$\alpha$), their quantum amplitudes are coherent. Therefore it is
important that \emph{the same} damping factor $\tilde\nu_\alpha$
be used in \emph{all} the  transitions (cf.\ the discussion
of radiative cascades in quantum oscillator by
\citealt{Cohen-T-ea}). Moreover, the same $\tilde\nu_\alpha$
given by \req{damp} should be used for the absorption and
scattering processes. This
ensures that the cyclotron
cross-section, being integrated across the resonance, provides
the correct value of the cyclotron oscillator strength (e.g.,
\citealp{Ventura79}), otherwise the
equivalent width of the cyclotron line would be overestimated.

In the electron resonance region, where $|\omega-\omce|\ll\omce$ and
$\alpha=-1$, one can neglect $\omcp/\omega$, 
because it is much smaller than 1, and the term that contains
$\nu_\mathrm{p}$, because it is small compared to the other terms.
The result coincides with the conventional
expression for the electron free-free cross-section without
allowance for proton motion with
$\tilde\nu_{-1}=\nu_{-1}^\mathrm{ff}+\nu_\mathrm{e}$. 
In the proton resonance region, where $|\omega-\omcp|\ll\omcp$ and
$\alpha=+1$, 
 the denominator in \req{sigma-fit} becomes
$
(\omega+\omce)^2 (\omega-\omcp)^2
             +\omega^2 \tilde\nu_\alpha^2
             \approx
             \omce^2\,[(\omega-\omcp)^2+\tilde\nu_\mathrm{p}^2],
$
{where}
$
\tilde\nu_\mathrm{p} = (\mel/\mpr)\,\nu_{\alpha}^{\mathrm{ff}}(\omcp).
$
In this approximation,
\req{sigma-fit} becomes
formally equivalent to a simple one-particle
cyclotron cross-section (cf.~Eq.~14 of \citealt{Pavlov95}
or Eq.~47 of \citealt{sawyer07}), apart from a difference 
in notations and
the difference in 
$\Lambda^\mathrm{ff}$
(the latter being discussed in
Sect.\,\ref{sect:omcp}).

The treatment that leads to \req{damp}
predicts a small shift in the
position of the resonance due to the damping.
This shift is unimportant
for applications and therefore neglected in \req{sigma-fit}.

\section{Cyclotron harmonics}
\label{sect:cycl}

In addition to the fundamental cyclotron resonances, the quantum
treatment of the free-free absorption identifies electron and
proton cyclotron harmonics at integer multiples of $\omce$ and
$\omcp$, respectively. They appear because of the increase in the
partial Coulomb logarithms $\Lambda_{n,N;\,n',N'}^{(\alpha)}
(\beta_\ast,\omega/\omega_\ast)$ 
at $\omega\to \omcp(N'-N)+\omce(n'-n)$.
Thus, $l$th electron cyclotron harmonics
(in addition to the fundamental at $\omega=\omce$) arises at
$\omega=(l+1)\,\omce$ due to the terms with $n'-n=l+1$,
and each $l$th proton cyclotron harmonics
(additional to the fundamental at $\omega=\omcp$)
is formed by the terms
with $N'-N=l+1$ in \req{nu-ff}. 
Unlike the classical electron and proton cyclotron resonances,
the quantum peaks of $\Lambda^\mathrm{ff}$
contribute to $\sigma_\alpha(\omega)$
at any polarization and are the same for $\alpha=+1$
and $-1$.

The relative strengths of the harmonics depend on the distribution
numbers $f^e_n$ and $f^p_N$. In this paper, we assume local
thermodynamic equilibrium (LTE) and thus use the Boltzmann
distributions, as in most of the previous papers (but see
\citealt{nagelventura83} and \citealt{PL07} for 
non-LTE effects on the electron and proton
cyclotron radiation rates, respectively).

We calculate free-free cross-sections in 
magnetized neutron-star atmospheres using
Eqs.~(\ref{sigmasum})--(\ref{nu-ff}). Examples of opacities
and/or spectra calculated with the use of these cross-sections can be
found, e.g., in 
\citet{PC03,PC04}, \citet{potekhinetal04},
\citet{hoetal08}, \citet{SPW09,SPW10}. 
In previous studies, various
additional simplifications have been made
in addition to the nonrelativistic, dipole, first Born
approximations described above for the free-free cross-sections. 
Below we assess
the applicability ranges of
these simplifications by comparing with
our more accurate results.

\subsection{Electron and muon cyclotron harmonics}

\subsubsection{Fixed scattering potential}
\label{sect:Minf}

In early works \citep[e.g.,][and references therein]{Mesz},
free-free (or bremsstrahlung) processes were treated assuming
scattering off a fixed Coulomb center, which is equivalent to 
the approximation of $\mpr\to\infty$, described in
Sect.\,\ref{sect:infimass}. In this approximation, one can set
$\omcp=0$ and explicitly perform the summation over $N'$ in
\req{sigmasum} using the identity $\sum_{N'=0}^\infty
I_{N'N}^2(t) = 1$. Taking damping (Sect.\,\ref{sect:damp}) into
account, we obtain
\bea&&
   \sigma_\alpha(\omega) = \frac{4\pi e^2
          }{ 
     \mel c} \,
   \frac{\nu_\alpha^\mathrm{ff}(\omega)
          }{
          (\omega+\alpha\omce)^2
          + (\nu_\mathrm{e}+\nu_\alpha^\mathrm{ff})^2
             }\,,
\label{sigma-e}
\\&&
   \nu_\alpha^\mathrm{ff} = 
        \frac{4}{3}\,\sqrt{\frac{2\pi}{\mel T}}\,
          \frac{n_\mathrm{e}\, e^4}{\hbar \omega}\,
 \Lambda_\alpha^\mathrm{ff}(\bete,\omega/\omce),
\\&&
   \Lambda_\alpha^\mathrm{ff}
       (\bete,\omega/\omce)
   =   \frac32  \sum_{n}f^e_n  \sum_{n'}
   \int_0^\infty \frac{\dd u}{u'}\,\mathrm{e}^{-\bete u^2/2}
   \,\theta({u'}^2)\,
\nonumber\\&&\qquad\times
    \big((u'+ u)^{2|\alpha|}\, w_{n;\,n'}^{(\alpha)}(u_+)
    + (u'- u)^{2|\alpha|}\, w_{n;\,n'}^{(\alpha)}(u_-)
    \big),
\label{Lambdann}
\\&&
       w_{n;\,n'}^{(\alpha)}(u_\pm)
    = \frac12 \int_0^\infty \frac{t^{|\alpha|}\,\dd t}{(t+u_\pm^2/2)^2}
   I_{n',n}^2(t),
\label{wnn}
\eea
where $\bete= \hbar\omce/T$
and
$
   {u'}^2 = u^2 + 2\,(n-n')
    + {2\omega}/\omce.
$
Assuming Boltzmann distribution 
($f^e_n/f^e_0=2\mathrm{e}^{-n\bete}$
at $n\geq1$, where the factor 2 
takes account of the electron spin degeneracy),
one can reduce this result to Eq.~(27) of \citet{PP76}
(as corrected by \citealt{PC03}).

\begin{figure}
\includegraphics[width=\columnwidth]{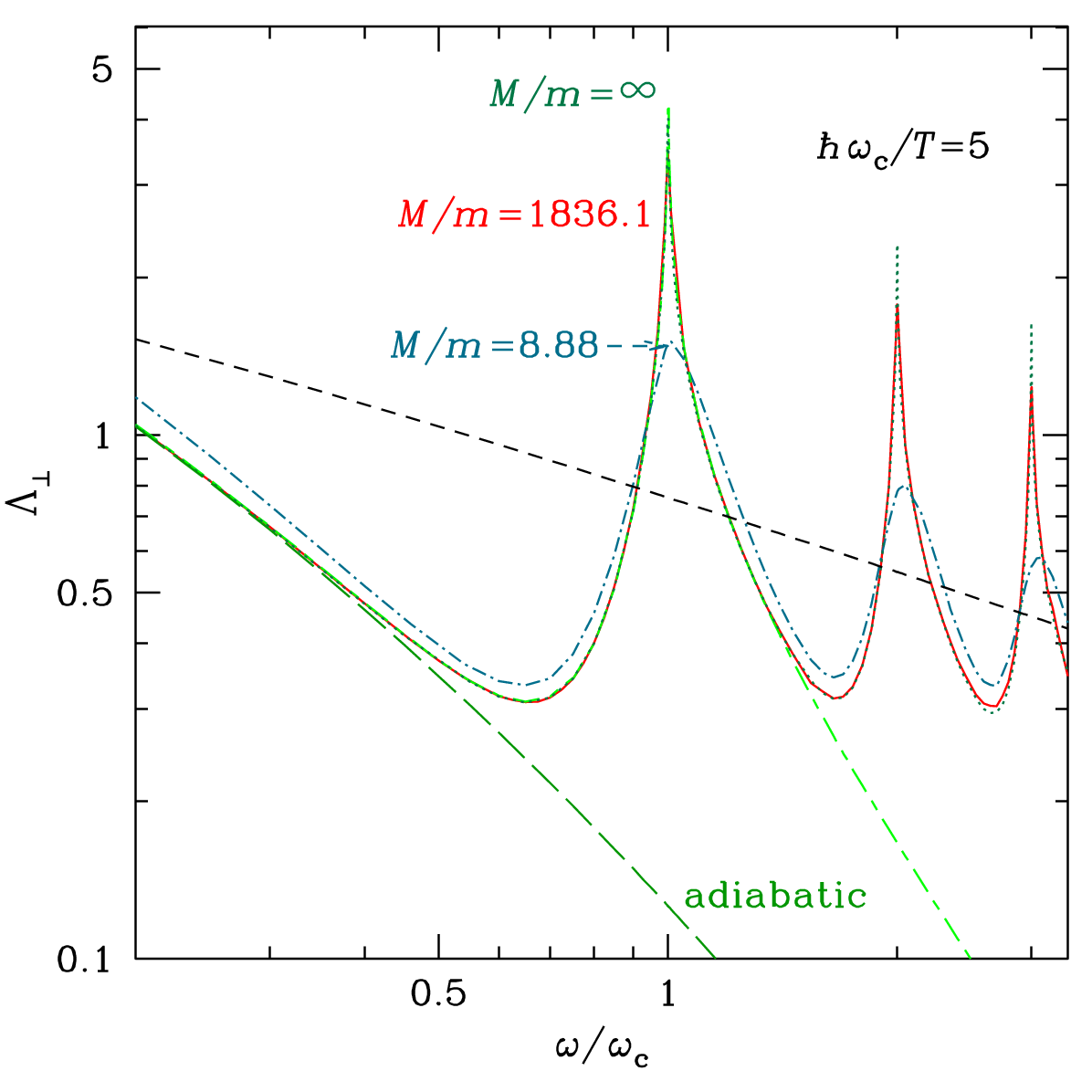}
\caption{
Transverse Coulomb logarithm as a function of $\omega/\omce$
at $\hbar\omce/T=5$
for different approximations: the model of fixed Coulomb potential
(dotted line),
approximate account of proton recoil
(solid line), 
adiabatic approximation (long-dashed line),
and the first post-adiabatic approximation (short dash - long
dash). The divergent peaks are trimmed at
$|\omega-(l+1)\,\omce|=10^{-3}\omce$ ($l=0,1,2,\ldots$).
For comparison, the nonmagnetic Coulomb logarithm 
(short dashes; in this case the horizontal axis 
displays $\hbar\omega/5T$)
and the model with approximate account of proton recoil
in the muonic atom $\mu^-p$ (dot-dashed line) 
are shown.
}
\label{fig:ql10mes1m}
\end{figure}
\begin{figure}
\includegraphics[width=\columnwidth]{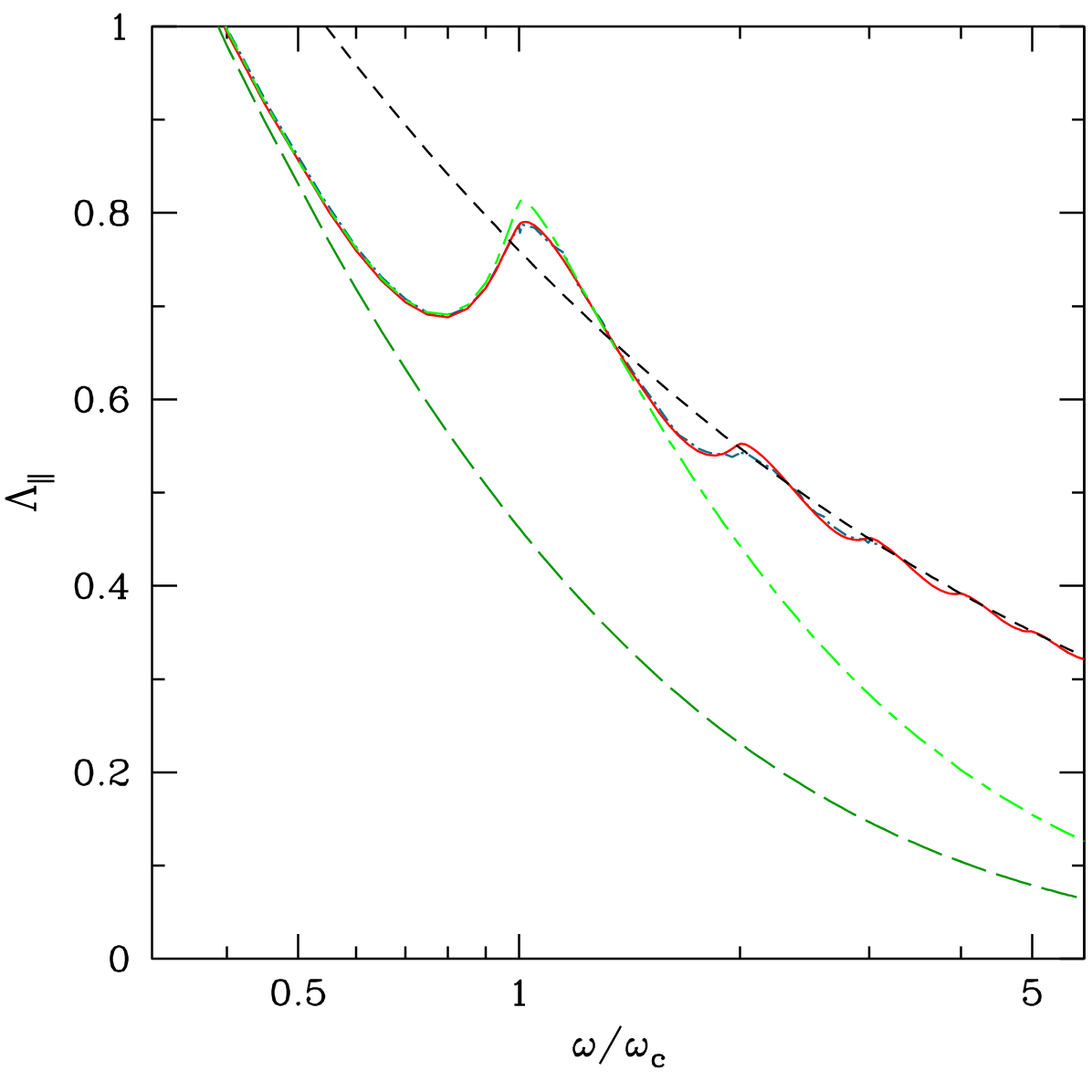}
\caption{
The same as in Fig.~\ref{fig:ql10mes1m}
but for the longitudinal Coulomb logarithm. 
In this case, lines with the approximate account of proton recoil
almost coincide with the line 
corresponding to the fixed-potential approximation.
}
\label{fig:ql10mes0m}
\end{figure}

This approximation was used in all models
of the spectra of strongly magnetized neutron stars
until the beginning of the 21st century
(e.g., \citealt{Pavlov95,zaneetal00}, and references therein).
It is validated by
the large value of the mass ratio $\mpr/\mel$.
In addition, it requires that $\omega\gg\omcp$,
as seen directly from the comparison of \req{sigma-e} with
\req{sigma-fit}.
   
\subsubsection{Approximate account of proton recoil}
\label{sect:recoil}

\citet{PP76} proposed
an approximate treatment of proton recoil, which assumes
that $\bete\ll\mpr/\mel$ and does not take into account Landau
quantization of proton motion.
In Fig.~\ref{fig:ql10mes1m}, the dotted line shows the
perpendicular Coulomb logarithm $\Lambda_\perp^\mathrm{ff}$
calculated according to Eqs.~(\ref{sigma-e})--(\ref{wnn}), while
the solid line takes the approximate account of proton recoil. 
As an example, we show the case where $\bete=5$. 
The familiar
nonmagnetic Coulomb logarithm in the first Born approximation
(e.g., \citealt{BetheSalpeter}) is shown by the short-dashed
line, assuming the same $\hbar\omega/T$
along the horizontal axis as for the other curves
($\hbar\omega/T=\bete\omega/\omce$).

To enhance the difference caused by the recoil,
we replace the electron by the muon $\mu^-$. All the
above formulae and discussion remain unchanged, but now the mass
ratio is $M/\mel=8.88$. The result of the 
approximate treatment of the recoil is shown by the dot-dashed
line.

In Fig.~\ref{fig:ql10mes0m}, the same approximations are shown
for $\Lambda_\|^\mathrm{ff}$. In
this case, the lines related to the cyclotron harmonics are 
smoothed, because the factors $(u'\pm u)^2$ quench the
near-threshold growth of the integrand in \req{Lambdann}. The
same smoothing results in the
infinite proton mass approximation being even more applicable
(under the necessary condition
$\omega\gg\omcp$): the dotted, solid, and dot-dashed lines almost
coincide in Fig.~\ref{fig:ql10mes0m}.

\subsubsection{Adiabatic and post-adiabatic approximations}
\label{sect:ad}

Several authors \citep{VirtamoJauho,nagelventura83,Mesz}
used the adiabatic approximation not only for the unperturbed
wave function $\psi^{(0)}$,
but also for $\psi^{(1)}$.
 This was done in addition to assuming the infinite proton
mass (Sect.\,\ref{sect:Minf}).
In other words, they kept only one $(n,s)$ term in the sum
given by \req{psi1e}. The result
is shown in Figs.~\ref{fig:ql10mes1m} and \ref{fig:ql10mes0m} by
long-dashed lines. We see that this approximation works well at
$\omega\ll\omce$, but becomes inaccurate at
$\omega\gtrsim\omce$.

\citet{sawyer07} analyzed the
photoabsorption problem
by using the method of field theory.
In the region $\omcp\ll\omega\leq1.5\omce$, he considered
account two electron Landau levels $n=0$ and 1
and applied a perturbation theory assuming the parameter
$\varepsilon=\mathrm{e}^{-\bete/2}$ to be small.
His result is identical to the results discussed in
Sect.\,\ref{sect:Minf} expanded in powers of $\varepsilon$,
which
we can write as
\bea
  \Lambda_\perp^\mathrm{ff} &=&
  \frac34\,\mathrm{e}^{\hbar\omega/T}
  \int_0^\infty \frac{\dd y}{(1+y)^2}\,
  \Big(\,K_0(x_0)
\nonumber\\&&
    + \,\frac{2\mathrm{e}^{-\bete/2}}{1+y}
   \big( K_0(x_1)+K_0(x_{-1})\,\big)\,\Big).
\label{Lambda1e}
\eea
Here and in the next equation, 
$K_\nu(x_n)$ are modified Bessel functions, and
$
   x_n\equiv |\hbar\omega/T+n\bete|\sqrt{0.25+y/\bete}.
$
Equation (\ref{Lambda1e}) differs 
from Eq.\,(28) of \citet{sawyer07}
in two respects: first, we have restored the factor 2
at $\mathrm{e}^{-\bete/2}$, and second,
we have dropped a term proportional to
$\mathrm{e}^{-\bete}$, 
because it is of the same order $\varepsilon^2$ as
the contribution from the level $n=2$, and therefore
should be treated together with the latter contribution
in the next order of the perturbation theory.

In the same way, we obtain
\bea
  \Lambda_\|^\mathrm{ff} &=&
  \frac34\,\mathrm{e}^{\hbar\omega/T}
  \int_0^\infty \frac{\dd y}{(1+y)\,(\bete/4+y)}\,
  \Big(\,x_0\,K_1(x_0)
\nonumber\\&&
    +\, \frac{\mathrm{e}^{-\bete/2}}{1+y}
   \big( x_1 K_1(x_1)+x_{-1} K_1(x_{-1})\,\big)\,\Big).
\label{Lambda0e}
\eea
Equations (\ref{Lambda1e}) and (\ref{Lambda0e})
can be obtained by the first iteration in 
the perturbation expansion,
starting from the adiabatic aproximation.

\subsection{Proton cyclotron harmonics}
\label{sect:omcp}

\begin{figure}
\includegraphics[width=\columnwidth]{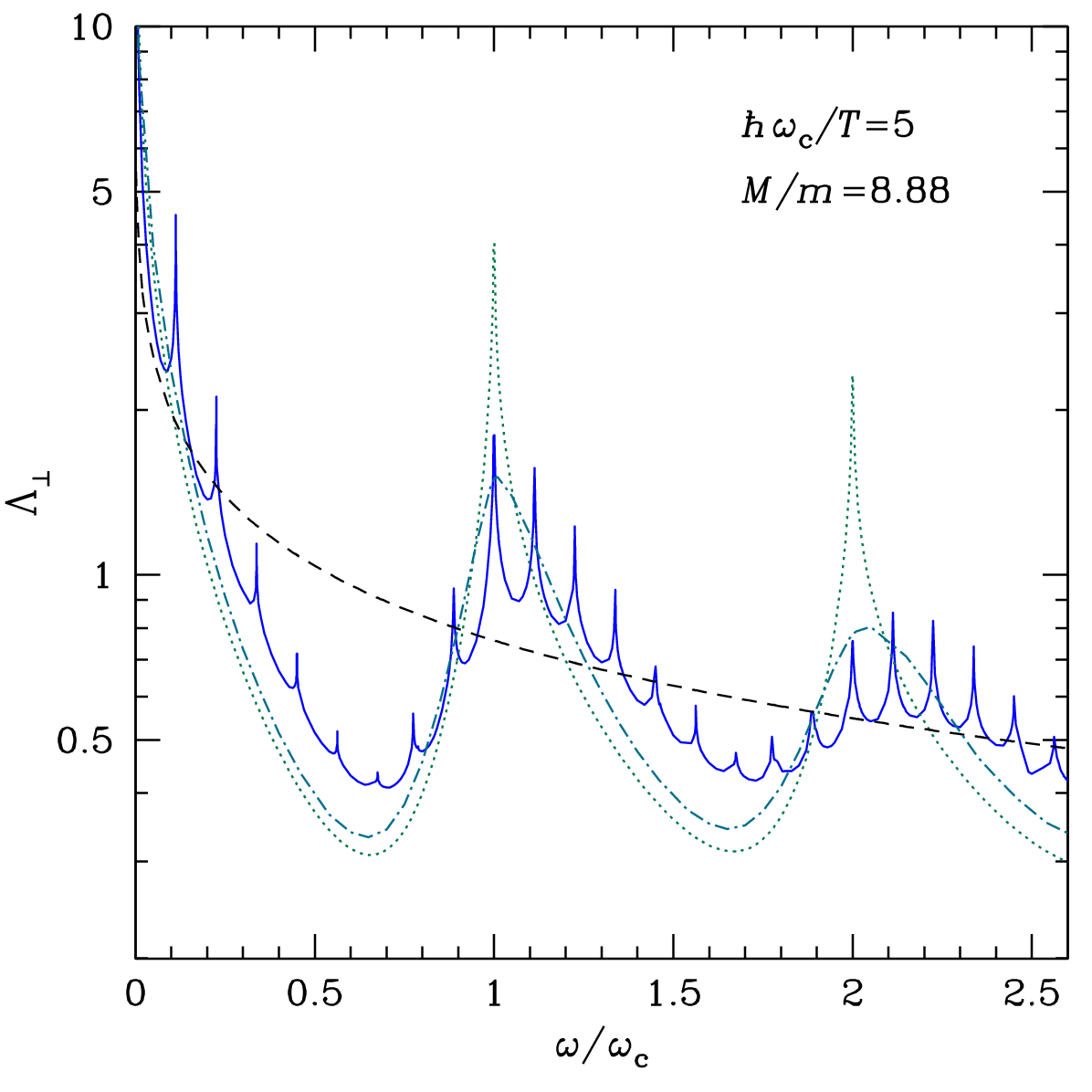}
\caption{
Transverse Coulomb logarithm as function of $\omega/\omce$
at $\hbar\omce/T=5$
in the model of fixed Coulomb potential
(dotted line) and the
approximate account of proton recoil
(dot-dashed line),
compared to the nonmagnetic Coulomb logarithm
(dashed line) and the accurate calculation
for the muonic atom $\mu^-p$ (solid line).}
\label{fig:ql10mes1}
\end{figure}

Proton cyclotron harmonics in the photoabsorption coefficients at
$\omega=(l+1)\,\omcp$ are superimposed on the peaks related to the
electron cyclotron harmonics. However, for the H atom the two
series of harmonics are separated because of the large value of
$\mpr/\mel=1836.1$. To observe the superimposition and the
qualitative differences of various approximations, it is
instructive to consider, in place of the H atom,
the muonic atom (the $\mu^-p$
system), which has a smaller mass ratio $M/\mel=8.88$. The
transverse Coulomb logarithm $\Lambda_\perp^\mathrm{ff}$ of
photoabsorption by such
system is shown in Fig.~\ref{fig:ql10mes1}. The solid line
displays the result of a calculation made according to 
Sect.\,\ref{sect:ff}. The other lines, as well as in
Fig.~\ref{fig:ql10mes1m}, show the results of different
approximations: a fixed Coulomb center (Sect.\,\ref{sect:Minf},
dotted line), the approximate account of  proton recoil
(Sect.\,\ref{sect:recoil}, dot-dashed line), and a nonmagnetic Coulomb
logarithm (dashes). 

The smaller peaks in the solid curve correspond to the proton
cyclotron harmonics. They are superimposed on the large-scale
oscillations, which correspond to the muon cyclotron
harmonics. Although the approximate recoil treatment (dot-dashed
line) improves the agreement with the exact calculation compared
to the infinite proton mass model (dotted line), both that approximate
models that neglect proton Landau quantization
differ significantly from the precise result.

\begin{figure}
\includegraphics[width=\columnwidth]{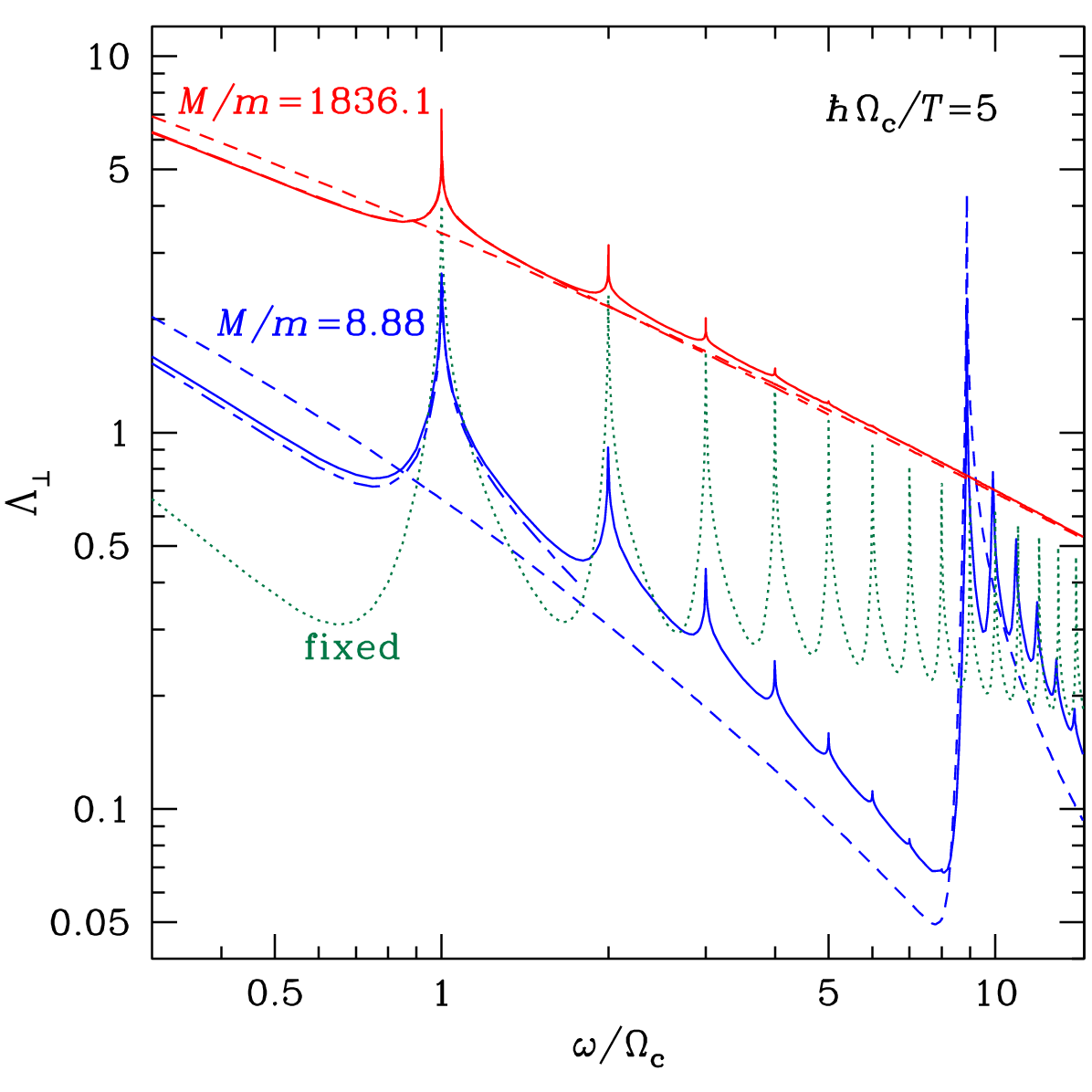}
\caption{
Transverse Coulomb logarithm as function of $\omega/\omcp$
at $\hbar\omcp/T=5$.
The accurate calculation (solid line)
for the systems $\mu^-p$ (lower lines) and $ep$ (upper lines)
is compared to the approximation of a fixed Coulomb
potential for the electron or muon scattering
(dashed lines) or for the proton scattering (dotted line),
and to the first post-adiabatic approximation (short dash - long
dash).}
\label{fig:ql10mes2p}
\end{figure}

In Fig.~\ref{fig:ql10mes2p}, we compare the proton cyclotron
harmonics for different relative masses of the positive and
negative particles. Here the
\emph{proton} cyclotron parameter is fixed to $\betp=\hbar\omcp/T=5$,
and the horizontal axis displays the ratio $\omega/\omcp$. The
solid lines show the transverse Coulomb logarithm for the muonic
atom (the lower curve) and the H atom (the upper curve). The
dashed lines show $\Lambda_\perp^\mathrm{ff}$ calculated
for the same $\hbar\omce/T=\betp\mpr/\mel$ 
and the same $\omega/\omce=(\mel/\mpr)\,\omega/\omcp$
as the solid curves,
but for the approximation of a fixed Coulomb potential
in the electron or muon scattering.
By comparison, the dotted line shows $\Lambda_\perp^\mathrm{ff}$
calculated for proton scattering off a fixed Coulomb
center, which can be regarded as a model where
$M/\mel\to0$. We see that  the approximate models are unable to
reproduce the proton cyclotron features correctly. It is also
noteworthy that the larger the ratio $M/\mel$, the smaller
the proton cyclotron peaks. In addition,
the cyclotron resonance strength 
decreases with increasing harmonics number $l$. 
These properties of the
cyclotron harmonics allow us to conclude that the solid lines in
Figs.~\ref{fig:ql10mes1m} and~\ref{fig:ql10mes0m} 
are precise (proton cyclotron harmonics are
negligible on their scale).

In the early models of magnetized neutron star atmospheres
\citep[e.g.,][]{pavlovshibanov78,shibanovetal92,SZ95}, the authors
considered moderate magnetic fields $B\sim10^{11}$--$10^{12}$~G,
where the proton Landau quantization is unimportant. More recently, 
observational evidence has accumulated
that some of the isolated neutron stars are probably magnetars,
which have fields of $B\sim10^{14}$~G (see, e.g., the review by
\citealt{Mereghetti08} and references therein). According to
\req{omcp}, the proton cyclotron lines of magnetars are in
an observationally accessible spectral range, which has
encouraged theoretical modeling of these features. In the absence
of an accurate quantum treatment, several authors
\citep{zaneetal00,zaneetal01,Ozel01,holai01,holai03} employed the
scaling previously suggested for this purpose by
\citet{Pavlov95}, according to which the
free-free cross-section for protons equals
$(\mel/\mpr)^2\,\sigma_{-\alpha}(\omega\mpr/\mel)$, where
$\sigma_\alpha(\omega)$ is given by \req{sigma-e}. 
The latter equation
differs remarkably from the correct expression in \req{sigma-fit}.
At photon frequencies $\omega<\omcp$, the difference roughly
amounts to a factor of $(\omega/\omcp)^2$.

In addition, the Coulomb logarithm that determines
$\nu_\alpha^\mathrm{ff}$ cannot be obtained from this scaling. 
An example is shown in
Fig.~\ref{fig:ql10mes2p}, where the dotted line corresponding to the
fixed-potential model is compared with the accurate calculations
displayed by the solid lines. We see that the fixed-potential
model strongly overestimates the strength of the proton cyclotron
harmonics. The origin of the discrepancy is clear: while
considering a collision of a proton with an electron, one cannot
assume the electron to be a nonmoving particle.

\citet{sawyer07} employed a representation with definite $s_\mathrm{e}$
and $s_\mathrm{p}$ and analyzed the first proton-cyclotron
peak of $\Lambda_\perp^\mathrm{ff}$, in a way similar to his analysis of
the first electron cyclotron peak (see Sect.\,\ref{sect:ad}),
by taking into account the ground electron Landau level $n=0$ and
two proton Landau levels, $N=0$ and 1. The result (his Eq.~30) is
quite accurate close to the fundamental cyclotron frequency, as we
illustrate wuth the lines of alternating short and long dashes in
Fig.~\ref{fig:ql10mes2p}. In the case of hydrogen (higher $M/m$),
it almost coincides with the accurate result (solid line) at
$\omega\lesssim1.5\,\omcp$ and with the result obtained by
neglecting the Landau quantization of protons (dashed line) at higher
$\omega$ values.

\section{Discussion} 
\label{sect:discussion}

\subsection{Corrections beyond Born approximation}

The formulae presented in Sects.\,\ref{sect:exact} and
\ref{sect:gen} in principle allow one to perform an accurate
calculation of photoabsorption rates in the electron-proton
system in an arbitrary magnetic field, taking into account the
effects of Landau quantization of the electron and proton motion
across the field and the transverse motion of the center of mass.
For bound-free absorption, this calculation was presented by
\citet{PP97}. For free-free processes, we apply the first Born
approximation and the dipole approximation. We plan to perform
calculations of the free-free opacities beyond
Born approximation in future work. 
An approximate estimate of the non-Born corrections
can be obtained \citep{PL07} by introducing
correction factors  $(1+\gamma_\ast^{-1}
u^{-2})^{-1/2}(1+\gamma_\ast^{-1} (u')^{-2})^{-1/2}$ into the
integral of \req{Lambda0}, where 
$\gamma_\ast=(\mel/\mred)^2\gamma$ and
$ \gamma = \hbar^3
B/(\mel^2 c e^3) = 425.44\,B_{12}.
$ 
The accuracy of the approximation
is ensured by the smallness of
$\gamma_\ast^{-1/2}\approx0.05\,B_{12}^{-1/2}$ 
and the additional condition $T\gg e^4 m_\ast/\hbar^2$, 
which is the usual applicability
condition for a Born approximation without a magnetic field. 

We have checked that these 
corrections are sufficiently small
for the electron cyclotron harmonics at $B\sim10^{11}$~G 
(relevant to CCOs) and negligible
 for the proton
cyclotron harmonics at $B>10^{13}$~G (relevant to XDINSs).

\subsection{Importance of bound states}

\begin{figure}
\includegraphics[width=\columnwidth]{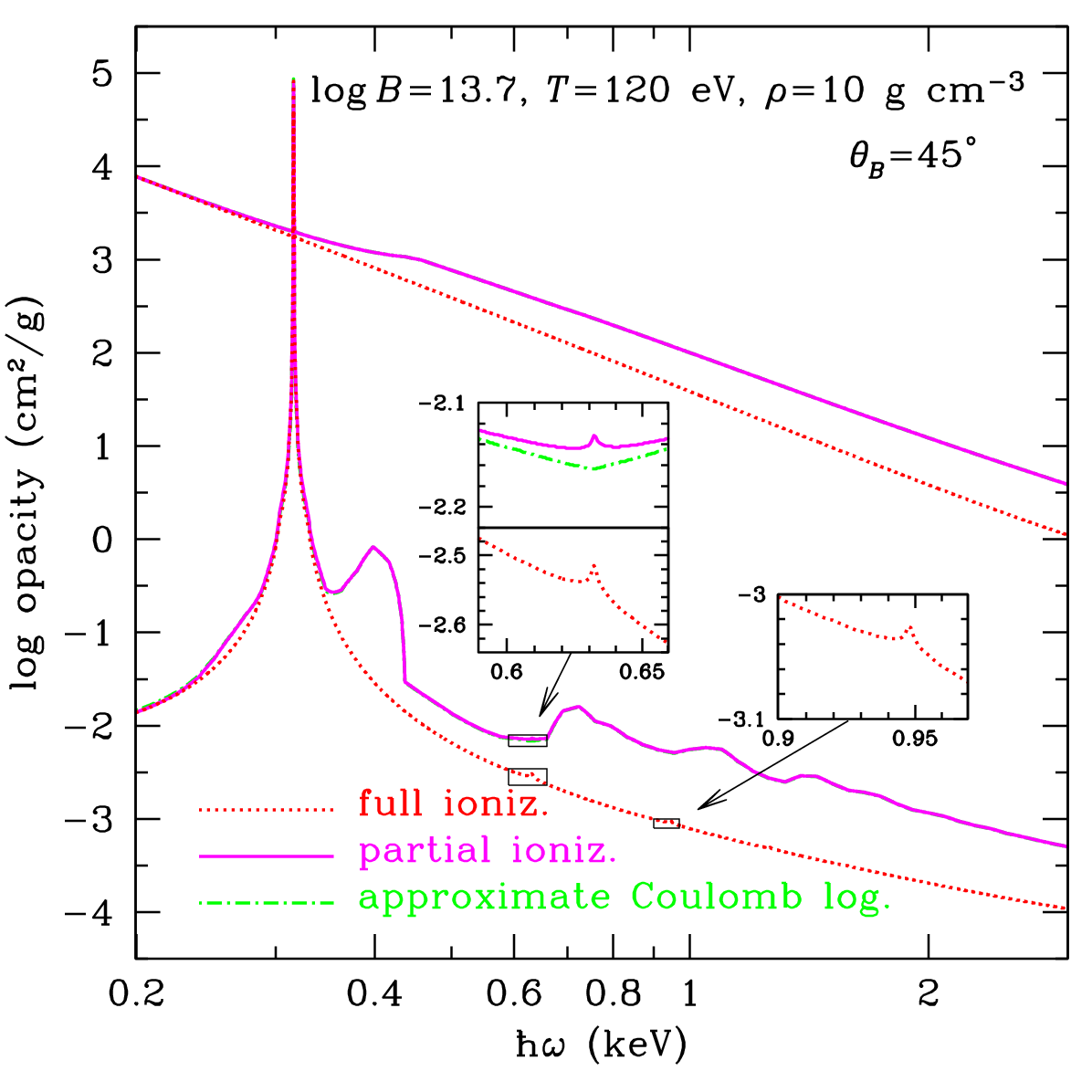}
\caption{
Opacities for the two normal electromagnetic waves propagating at the angle
$\theta_B=45^{\circ}$ to the magnetic field direction in a
hydrogen atmosphere of a neutron star with $B=5\times10^{13}$~G
and $T=120$~eV at density $\rho=10$~\gcc\ (which is in the
middle of the photosphere at these $B$ and $T$). The results are
shown for fully ionized (dotted lines) and partially ionized
(solid and dot-dashed lines) 
plasma models.
In the latter model, the nonionized atomic fraction equals 0.0066.
The solid line shows the
opacity obtained with the accurate calculation of the free-free
Coulomb logarithm, and the dot-dashed line demonstrates the result of
the approximate treatment that corresponds to the dashed line
in Fig.~\ref{fig:ql10mes2p}. }
\label{fig:ang13_7c2}
\end{figure}

Free-free absorption contributes only a part of
the total opacities in the atmospheres of neutron stars. A
second constituent is the familiar scattering, and a third the absorption
by bound species \citep[see, e.g.,][]{CanutoVentura,Pavlov95}. 
It was realized long ago \citep{Ruderman71}
that in strong magnetic fields the increase in the binding energies of atoms and molecules
can lead to their non-negligible
abundance even in hot atmospheres. With increasing $B$, the
binding energies and abundances of bound species increase at any
fixed density $\rho$ and temperature $T$ \citep{PCS99,Lai01}, so that
even the lightest of the atoms, hydrogen, provides a noticeable
contribution to the opacities at the temperatures of interest,
if the magnetic field is strong enough. 
Even a small neutral fraction can be important, 
because the bound-bound and bound-free
cross-sections are large close to certain characteristic
spectral energies. 

For electron cyclotron harmonics to appear at
$\hbar\omega\lesssim1$ keV,
we should ensure that $B\lesssim10^{11}$~G according to \req{omce}. At
these relatively weak magnetic fields and the characteristic
temperature $T^\infty\gtrsim100$ eV, the assumption of full
ionization may be acceptable. However, at $B>10^{13}$~G, 
which is required for ion cyclotron harmonics,
the situation is different. An illustration is given in
Fig.~\ref{fig:ang13_7c2}.
The solid curves show true absorption
opacities for two normal electromagnetic waves propagating at
the angle $45^\circ$ to the magnetic field lines
at $B=5\times10^{13}$~G and $T=120$~eV.
The upper and
lower curves correspond to the ordinary and extraordinary waves,
respectively.
The density in this example is chosen to be $\rho=10$~\gcc, which is
a typical atmosphere density at $B=5\times10^{13}$~G and
$\Teff^\infty=100$ eV (at this
density the thermodynamic temperature $T$ approximately equals
the effective temperature $\Teff$). According to our ionization
equilibrium model \citep{PCS99}, at these $B$, $T$, and $\rho$
values, 0.66\% of protons in the plasma are comprised in the
ground-state H atoms that are not too strongly perturbed by
plasma microfields so that they contribute to the bound-bound and
bound-free opacities (the ``optical'' atomic fraction), and only
0.1\% of protons are in excited bound states. Even though the
ground-state atomic fraction is small, it is not negligible. In
Fig.~\ref{fig:ang13_7c2}, at $\hbar\omega\gtrsim0.4$ keV, the
opacities in two normal modes, calculated with allowance for 
partial ionization (solid and dot-dashed curves), are significantly
higher and have more characteristic features than the opacity
calculated under the assumption of complete ionization
(dotted lines). In particular, the broad feature on the lower
curve near 0.4 keV is
produced by the principal bound-bound transition between the two
lowest bound states ($s_i=0\to s_f=1$), 
and the increased value of the opacity at
higher energies $\hbar\omega$ is due to the transitions to
other bound and free quantum states. 
The wavy shape of the lower solid curve (for the
extraordinary mode) 
at $\hbar\omega\gtrsim0.7$
keV is explained by
bound-free transitions to different open channels, each having
its own threshold energy.
All the bound-bound absorption features
and photoionization thresholds are strongly broadened by
the effects of atomic motion across the magnetic field lines
(``magnetic broadening'', see \citealt{PP97} and references
therein).

In the insets, we zoom in on the regions of the first and second
proton cyclotron harmonics. Both of them are visible, but
negligible compared to the effect of partial ionization on the
opacities.

\subsection{Other possibilities for CCOs and XDINSs}

Apart from the cyclotron harmonics, 
a number of alternative
explanations of the observed absorption features in CCOs and
XDINSs have been suggested in the literature.

\citet{moriho07} constructed models of strongly magnetized neutron
star atmospheres with mid-$Z$ elements
and compared them to the observed spectra of the
neutron stars 1E 1207.4-5209 and RX J1605.3+3249. They demonstrated
that the positions and relative strengths of the strongest absorption
features in these neutron stars are in good agreement with a
model of a strongly ionized oxygen atmosphere with $B=10^{12}$~G
and  $B=10^{13}$~G, respectively. This explanation seems
promising, but unsolved problems remain: the effects of
motion across the field have been treated approximately,
based on the assumption that they are small, and detailed fits to
the observed spectra have not yet been presented.

Among other hypotheses about the nature of the absorption features,
there was a suggestion that they
could be due to bound-bound 
transitions in exotic molecular ions \citep{TurbinerLopez06}.
However, 
our estimates show that the abundance of these ions
in a neutron star atmosphere
would be negligible compared with the
abundance of H atoms.
\citet{SPW09} proposed a ``sandwich'' model atmosphere of finite
depth, composed of a helium slab above a condensed surface and
beneath hydrogen, and demonstrated
that this model can produce
two or three absorption features in the range
of $\hbar\omega\sim0.2$--1 keV at $B\sim10^{14}$~G, although 
a detailed
comparison with observed spectra was not performed.
One cannot also rule out that some
absorption
lines originate in a
cloud near a neutron star, rather than in the atmosphere
\citep[see][]{hambaryanetal09}.

\section{Summary} 
\label{sect:sum}

We have considered the basic methods for calculation of
free-free opacities of a magnetized hydrogen plasma. Our
emphasis has been on the case where not only electron, but also proton
motion across the magnetic field is quantized by the Landau
states. We have derived general formulae for
the photoabsorption rates
and considered in detail the dipole,
first Born approximation. We have presented numerical examples,
compared them with the results of previously used simplified
models, and analyzed the physical assumptions behind the different
simplifications and conditions 
of their applicability. We have demonstrated that the
proton cyclotron harmonics at a given value of the parameter
$\betp=\hbar\omcp/T$ are much weaker than the respective
electron cyclotron harmonics at the same value of
$\bete=\hbar\omce/T$, and explained this difference in terms the large
(nonperturbative) effects of proton motion in the
case of proton cyclotron harmonics, in contrast to the case
of the electron cyclotron harmonics.

\begin{acknowledgements}
I am pleased to acknowledge
enlightening discussions with Gilles Chabrier, G\'erard Massacrier,
Yura Shibanov, and Dima Yakovlev, and 
useful communications with Ray Sawyer.
This work is partially supported by the RFBR Grant 08-02-00837
and Rosnauka Grant NSh-3769.2010.2.
\end{acknowledgements}
 


\begin{thebibliography}{88}


\bibitem[Araya \& Harding(1999)]{ArayaHarding}
Araya, R.~A., \& Harding, A.~K.
 1999, 
\apj, {517}, 334

\bibitem[Araya-G\'ochez \& Harding(2000)]{ArayaG-Harding}
Araya-G\'ochez, R.~A., \& Harding, A.~K.
 2000,
 \apj, {544}, 1067

\bibitem[Armstrong \& Nicholls(1972)]{Armstrong}
Armstrong, B.~M., \& Nicholls, R.~W.
 1972,
{Emission, Absorption and Transfer of Radiation in Heated Atmospheres}
(Pergamon, Oxford)

\bibitem[Bekefi(1966)]{Bekefi}
Bekefi, G.
1966,
{Radiation Processes in Plasmas}
(Wiley, New York)

\bibitem[Bethe \& Salpeter(1957)]{BetheSalpeter}
Bethe, H.~A., \& Salpeter, E.~E.
 1957,
{Quantum Mechanics of One- and Two-Electron Atoms}
(Springer, Berlin)

\bibitem[Bezchastnov \& Potekhin(1994)]{BP94}
Bezchastnov, V.~G., \& Potekhin, A.~Y.
 1994,
\JPB{27}, 3349

\bibitem[Bignami et al.(2003)]{bignamietal03}
Bignami, G.~F., Caraveo, P.~A., De Luca, A., \& Mereghetti, S.
2003,
\nat, {423}, 725

\bibitem[Canuto \& Ventura(1977)]{CanutoVentura}
Canuto, V., \& Ventura, J.
1977,
\FCP{2}, 203

\bibitem[Cohen-Tannoudji et al.(1998)Cohen-Tannoudji, Dupont-Roc, \& Grynberg]{Cohen-T-ea}
Cohen-Tannoudji, C., Dupont-Roc, J., \& Grynberg, G.
1998,
{Atom-Photon Interactions:
Basic Processes and Applications}
(Wiley, Berlin)

\bibitem[Cropper et al.(2007)]{cropperetal07}
Cropper, M., Zane, S., Turolla, R., et al. 
2007,
\apss, {308}, 161

\bibitem[Daugherty \& Ventura(1977)]{DV77}
Daugherty, J.~K., \& Ventura, J.
 1977,
\aap, {61}, 723

\bibitem[de Luca(2008)]{deluca08}
 de Luca, A.
 2008,
AIP Conf.\ Proc., {983}, 311

\bibitem[de Luca et al.(2004)]{delucaetal04}
 de Luca, A., Mereghetti, S., Caraveo, P.~A., et al. 
 2004,
 \aap, {418}, 625
 
\bibitem[Enoto et al.(2008)]{enotoetal08}
Enoto, T., Makishima, K., Terada, Y., et al.
2008,
\pasj, 60, S57

\bibitem[Ginzburg(1970)]{Ginzburg}
Ginzburg, V.~L.
 1970,
{The Propagation of Electromagnetic Waves in Plasmas},
2nd ed. 
(Pergamon, London)

\bibitem[Gnedin \& Syunyaev(1974)]{GnedinSyunyaev74}
Gnedin, Yu.~N., \& Sunyaev, R.~A. 1974,
\aap, {36}, 379

\bibitem[Gor'kov \& Dzyaloshinskii(1968)]{GorDzyal}
Gor'kov, L.~P., \& Dzyaloshinskii, I.~E. 
1968,
\SovJETP{26}, 449

\bibitem[Haberl(2007)]{haberl07}
 Haberl, F.
  2007, 
 \apss, {308}, 181
 
\bibitem[Haberl et al.(2004)]{haberletal04}
 Haberl, F., Zavlin, V. E., Tr\"umper, J., \& Burwitz,~V.
 2004,
\aap, {419}, 1077
 
\bibitem[Halpern \& Gotthelf(2010)]{halperngotthelf10}
Halpern, J.~P., \& Gotthelf, E.~V.
 2010, 
\apj, {709}, 436

\bibitem[Hambaryan et al.(2009)]{hambaryanetal09}
Hambaryan, V., Neuh\"auser, R., Haberl, F., Hohle, M. M., \& Schwope, A. D.
2009,
\aap, {497}, L9

\bibitem[Ho \& Lai(2001)]{holai01}
 Ho, W.~C.~G. \& Lai, D.
  2001,
 \mnras, {327}, 1081

\bibitem[Ho \& Lai(2003)]{holai03}
 Ho, W.~C.~G. \& Lai, D.
  2003, 
 \mnras, {338}, 233
 
\bibitem[Ho et al.(2008)Ho, Potekhin, \& Chabrier]{hoetal08}
 Ho, W.~C.~G., Potekhin A.~Y., \& Chabrier, G. 
 2008, 
 \apjs, {178}, 102

\bibitem[Hohle et al.(2009)]{hohleetal09}
Hohle, M.~M.,  Haberl, F., Vink, J., et al. 
2009,
\aap, {498}, 811

\bibitem[Johnson \& Lippmann(1949)]{johnsonlippmann49}
Johnson, M.~H., \& Lippmann, B.~A.
1949,
\PRv{76}, 828

\bibitem[Johnson et al.(1983)Johnson, Hirschfelder, \& Yang]{johnsonetal83}
Johnson, B.~R., Hirschfelder, J.~O., \& Yang, K.-H.
1983,
\RMP{55}, 109

\bibitem[Kaplan \& van Kerkwijk(2009)]{kaplanvankerkwijk09}
Kaplan, D.~L., \& van Kerkwijk, M.~H.
2009,
\apj, {692}, L62
 
\bibitem[Kaspi et al.(2006)Kaspi, Roberts, \& Harding]{kaspietal06}
Kaspi, V.~M., Roberts, M.~S.~E., \& Harding, A.~K.
2006,
in {Compact Stellar X-Ray Sources}, ed.\ W.~Lewin \& M.~van der Klis
(Cambridge University Press, Cambridge, UK), 
279

\bibitem[Kopidakis et al.(1996)Kopidakis, Ventura, \& Herold]{KVH}
Kopidakis, N., Ventura, J., \& Herold, H.
 1996,
\aap, {308}, 747

\bibitem[Lai(2001)]{Lai01}
Lai, D.,
2001,
\RMP{73}, 629

\bibitem[Landau \& Lifshitz(1976)]{LaLi-QM}
Landau, L.~D., \& Lifshitz, E.~M.
 1976,
{Quantum Mechanics}
(Pergamon, Oxford)

\bibitem[Mereghetti(2008)]{Mereghetti08}
Mereghetti, S.
2008,
\aapr, {15}, 225

\bibitem[M\'esz\'aros(1992)]{Mesz}
M\'esz\'aros, P.,
 1992,
{High-Energy Radiation from Magnetized Neutron Stars}
(Chicago: Univ.~of Chicago Press)

\bibitem[Mori \& Ho(2007)]{moriho07}
 Mori, K. \& Ho, W.~C.~G. 2007, 
 \mnras, {377}, 905

\bibitem[Mori et al.(2005)Mori, Chonko, \& Hailey]{morietal05}
 Mori, K., Chonko, J.~C., \& Hailey, C.~J.
 2005,
 \apj, {631}, 1082
 
\bibitem[Nagel \& Ventura(1983)]{nagelventura83}
Nagel, W., \& Ventura, J.
 1983,
\aap, {118}, 66

\bibitem[Nishimura(2005)]{nishimura05}
Nishimura, O.
2005,
\pasj, {57}, 769

\bibitem[Nishimura(2008)]{nishimura08}
Nishimura, O.
2008,
\apj, {672}, 1127

\bibitem[{\"O}zel(2001)]{Ozel01}
{\"O}zel, F.
 2001,
\apj, {563}, 276

\bibitem[Pavlov \& Panov(1976)]{PP76}
Pavlov, G.~G., \& Panov, A.~N.
1976,
\SovJETP{44}, 300

\bibitem[Pavlov \& Shibanov(1978)]{pavlovshibanov78}
Pavlov, G.~G., \& Shibanov, Yu.~A.
1978,
\sovast, {22}, 214

\bibitem[Pavlov et al.(1980)Pavlov, Shibanov, \& Yakovlev]{PSY80}
Pavlov, G.~G., Shibanov, Yu.~A., \& Yakovlev, D.G.
1980,
\apss, {73}, 33

\bibitem[Pavlov et al.(1995)]{Pavlov95}
Pavlov, G.~G., Shibanov, Yu.~A., Zavlin, V.~E., \& Meyer, R.~D.
 1995,
in
{The Lives of the Neutron Stars}, NATO ASI Ser. C, {450},
ed.\ M.~A. Alpar, \"U. Kizilo\u{g}lu, \& J. van Paradijs
       (Kluwer, Dordrecht),
        71

\bibitem[Potekhin(1994)]{P94}
Potekhin, A.~Y.
 1994,
\JPB{27}, 1073

\bibitem[Potekhin(1998)]{P98}
Potekhin, A.~Y.
1998,
\JPB{31}, 49

\bibitem[Potekhin \& Chabrier(2003)]{PC03}
Potekhin, A.~Y., \& Chabrier, G.
 2003,
\apj, {585}, 955

\bibitem[Potekhin \& Chabrier(2004)]{PC04}
Potekhin, A.~Y., \& Chabrier, G.
2004,
\apj, {600}, 317

\bibitem[Potekhin \& Lai(2007)]{PL07}
Potekhin, A.~Y., \& Lai, D.
 2007,
\mnras, {376}, 793

\bibitem[Potekhin \& Pavlov(1997)]{PP97}
Potekhin, A.~Y., \& Pavlov, G.~G.
 1997,
\apj, {483}, 414

\bibitem[Potekhin et al.(1999)Potekhin, Chabrier, \& Shibanov]{PCS99}
 Potekhin, A.~Y., Chabrier, G., \& Shibanov, Yu.~A.
  1999, 
\pre, {60}, 2193; 
 erratum: \pre, {63}, 019901 (2000)

\bibitem[Potekhin et al.(2004)]{potekhinetal04}
 Potekhin, A.~Y., Lai, D., Chabrier, G., \& Ho, W.~C.~G.
  2004, 
 \apj, {612}, 1034
 
\bibitem[Pottschmidt et al.(2004)]{pottschmidtetal04}
Pottschmidt, K., Kreykenbohm, I., Wilms, J., et al.
\apj, {634}. L97
 
\bibitem[Rodes-Roca et al.(2009)]{rodes-rocaetal09}
Rodes-Roca, J.~J., Torrej\'on, J.~M., Kreykenbohm, I., et al. 
2009,
\aap, {508}, 395

\bibitem[Ruderman(1971)]{Ruderman71}
Ruderman, M.~A.
1971,
\prl, {27}, 1306

\bibitem[Santangelo et al.(1999)]{santangeloetal99}
Santangelo, A., Segreto, A., Giarusso, F., et al.
1999,
\apj, {523}, L85

\bibitem[Sanwal et al.(2002)]{sanwaletal02}
Sanwal, D., Pavlov, G.~G., Zavlin, V.~E., \& Teter, M.~A.
 2002,
\apj, {574}, L61 

\bibitem[Sawyer(2007)]{sawyer07}
Sawyer, R.~F.
2007,
arXiv:0708.3049v2 [astro-ph]

\bibitem[Schwope et al.(2007)]{schwopeetal07}
Schwope, A.~D., Hambaryan, V., Haberl, F., \& Motch, C.
2007,
\apss, {308}, 619

\bibitem[Schwope et al.(2009)]{schwopeetal09}
Schwope, A.~D., Erben, T., Kohnert, J., et al.
2009, 
\aap, {499}, 267

\bibitem[Seaton(1983)]{Seaton83}
Seaton, M.~J.
 1983,
\RPP{46}, 167

\bibitem[Shibanov \& Zavlin(1995)]{SZ95}
Shibanov, Yu.~A., \& Zavlin, V.~E.
 1995,
\AstL{21}, 3

\bibitem[Shibanov et al.(1992)]{shibanovetal92}
 Shibanov, Yu.~A., Zavlin, V.~E., Pavlov, G.~G., \& Ventura, J.
  1992,
 \aap, {266}, 313

\bibitem[Sokolov \& Ternov(1986)]{SokTer}
Sokolov, A.~A., \& Ternov, I.~M.
 1986,
Radiation from Relativistic Electrons,
2d ed. (AIP, New York)

\bibitem[Suleimanov et al.(2009)Suleimanov, Potekhin, \& Werner]{SPW09}
Suleimanov, V.~F., Potekhin, A.~Y., \& Werner, K.
2009,
\aap, {500}, 891

\bibitem[Suleimanov et al.(2010a)Suleimanov, Potekhin, \& Werner]{SPW10}
Suleimanov, V.~F., Potekhin, A.~Y., \& Werner, K.
2010a, 
\ASR{45}, 92

\bibitem[Suleimanov et al.(2010b)Suleimanov, Pavlov, \& Werner]{SulPavWer}
Suleimanov, V.~F., Pavlov, G.~G., \& Werner, K.
2010b, 
\apj, {714}, 630

\bibitem[Tr\"umper et al.(1978)]{Truemperetal78}
Tr\"umper, J., Pietsch, W., Reppin, C., et al.
1978,
\apj, {219}, L105


\bibitem[Turbiner \& L\'opez Vieyra(2006)]{TurbinerLopez06}
Turbiner, A.~V., \& L\'opez Vieyra, J.~C.
2006,
\physrep, {424}, 309

\bibitem[Turolla(2009)]{turolla09}
Turolla, R.
 2009,
in {Neutron Stars
 and Pulsars}, ed.\ W.~Becker, 
 Astrophys.\ Space Sci.\ Library, {357} 
(Springer, Berlin),
  141

\bibitem[van Kerkwijk(2004)]{vankerkwijk04}
 van Kerkwijk, M.~H.
 2004, 
in {Young Neutron Stars and Their Environments},
 Proceedings of IAU Symposium no. 218,
ed.\ F.~Camilo \& B.~M. Gaensler
(ASP, San Francisco),
283

\bibitem[van Kerkwijk \& Kaplan(2007)]{vankerkwijkkaplan07}
 van Kerkwijk, M.~H. \& Kaplan, D.~L.
  2007, 
\apss, {308}, 191

\bibitem[van Kerkwijk et al.(2004)]{vankerkwijketal04}
 van Kerkwijk, M.~H., Kaplan, D.~L., Durant, M., Kulkarni, S.~R.,
\& Paerels, F.
2004,
\apj, {608}, 432

\bibitem[Ventura(1979)]{Ventura79}
Ventura, J.
 1979,
\prd, {19}, 1684

\bibitem[Vincke \& Baye(1988)]{VB88}
Vincke, M., \& Baye, D.
1988,
\JPB{21}, 2407

\bibitem[Vincke et al.(1992)Vincke, Le Dourneuf, \& Baye]{VDB92}
Vincke, M., Le Dourneuf, M., \& Baye, D.
1992,
\JPB{25}, 2787

\bibitem[Virtamo \& Jauho(1975)]{VirtamoJauho}
Virtamo, J., \& Jauho, P.
1975,
\NuCi{26B}, 537

\bibitem[Wang et al.(1993)Wang, Wasserman \& Lamb]{WWL93}
Wang, J.~C.~L., Wasserman, I., \& Lamb, D.~Q.
 1993, 
\apj, {414}, 815

\bibitem[Wunner et al.(1983)]{Wunner-ea83}
Wunner, G., Ruder, H., Herold, H., \& Schmitt, W.
 1983, 
\aap, {117}, 156

\bibitem[Zane et al.(2000)Zane, Turolla, \& Treves]{zaneetal00}
 Zane, S., Turolla, R., \& Treves, A.
 2000, 
 \apj, {537}, 387

\bibitem[Zane et al.(2001)]{zaneetal01}
 Zane, S., Turolla, R., Stella, L., \& Treves, A.
  2001, 
 \apj, {560}, 384

\bibitem[Zavlin(2009)]{zavlin09}
 Zavlin, V.~E.
  2009, 
in {Neutron Stars
 and Pulsars}, ed.\ W.~Becker, 
 Astrophys.\ Space Sci.\ Library, {357} 
(Springer, Berlin),
  181

\bibitem[Zavlin et al.(1998)Zavlin, Pavlov, \& Tr\"umper]{zavlinetal98}
Zavlin, V.~E., Pavlov, G.~G., \& Tr\"umper, J.
 1998, 
\aap, {331}, 821

\end{thebibliography}
\end{document}